\documentclass[manuscript,screen]{acmart}

\AtBeginDocument{%
  }

\setcopyright{acmlicensed}
\copyrightyear{2025}
\acmYear{2025}
\acmDOI{XXXXXXX.XXXXXXX}

\acmISBN{978-1-4503-XXXX-X/2025/02}
\usepackage{xcolor}
\newcommand{\gray}{\cellcolor{gray!20}}
\usepackage{wrapfig}
\usepackage{colortbl}
\usepackage{multirow}
\usepackage{booktabs}
\usepackage[most]{tcolorbox}
\definecolor{HMTblue}   {HTML}{2F6BED}
\definecolor{HMTpurple} {HTML}{9A6BEE}
\definecolor{HMTpink}   {HTML}{E94E77}
\definecolor{HMTred}    {HTML}{F25A4E}
\definecolor{HMTorange} {HTML}{F5A623}
\definecolor{HMTamber}  {HTML}{F5C342}
\definecolor{HMTgreen}  {HTML}{7AC943}

\newtcolorbox{HMTcard}[2][]{%
  enhanced,
  boxrule=0pt,
  frame empty,
  colback=white,
  arc=2pt,
  left=2mm, right=2mm, top=1.2mm, bottom=1.2mm,
  boxsep=0.6mm,
  fonttitle=\bfseries\footnotesize,
  coltitle=white,
  title={#2},
  attach boxed title to top left={yshift=-1.0mm, xshift=1.2mm},
  boxed title style={
    boxrule=0pt,
    colback=#1,
    arc=5pt,
    top=0.8mm, bottom=0.8mm,
    left=2.5mm, right=2.5mm
  },
}

\newcommand{\nd}{\textcolor{black}}
\newcommand{\new}{\textcolor{black}}
\newcommand{\cbox}[1]{\raisebox{\depth}
{\fcolorbox{black}{#1}{\null}}}
\newcommand{\ctriangle}[1]{\raisebox{0.2ex}{\textcolor{#1}{$\blacktriangle$}}}
\newcolumntype{P}[1]{>{\centering\arraybackslash}p{#1}}
\usepackage{wrapfig}
\usepackage{tikz}
\usetikzlibrary{positioning, arrows.meta}
\usepackage{pgfplots}
\pgfplotsset{compat=1.16}

\begin{document}

\title{Toward Integrated Solutions: A Systematic Interdisciplinary Review of Cybergrooming Research}

% \author{Heajun An}
% \email{heajun@vt.edu}
% \affiliation{%
\author{Heajun An}
\email{heajun@vt.edu}
\orcid{0009-0007-6124-3750}
\affiliation{%
  \institution{Virginia Tech}
  \city{Blacksburg}
  \state{Virginia}
  \country{USA}}

\author{Marcos Silva}
\email{marcos.silva@aluno.cefetmg.br}
\orcid{0009-0006-9420-9535}
\affiliation{%
  \institution{Federal Center for Technological Education of Minas Gerais}
  \city{Belo Horizonte}
  \state{Minas Gerais}
  \country{Brazil}
}

\author{Qi Zhang}
\email{qiz21@vt.edu}
\orcid{0000-0002-3607-3258}
\affiliation{%
  \institution{Virginia Tech}
  \city{Falls Church}
  \state{Virginia}
  \country{USA}}

\author{Arav Singh}
\email{aravsingh@vt.edu}
\orcid{0009-0007-3582-9600}
\affiliation{%
  \institution{Virginia Tech}
  \city{Blacksburg}
  \state{Virginia}
  \country{USA}}

\author{Minqian Liu}
\email{minqianliu@vt.edu}
\orcid{0009-0001-6014-3949}
\affiliation{%
  \institution{Virginia Tech}
  \city{Blacksburg}
  \state{Virginia}
  \country{USA}}

\author{Xinyi Zhang}
\email{xinyizhang@vt.edu}
\orcid{0009-0003-4188-0104}
\affiliation{%
  \institution{Virginia Tech}
  \city{Blacksburg}
  \state{Virginia}
  \country{USA}}

\author{Sarvech Qadir}
\email{sarvech.qadir@vanderbilt.edu}
\orcid{0009-0006-7962-5792}
\affiliation{%
  \institution{Vanderbilt University}
  \city{Nashville}
  \state{Tennessee}
  \country{USA}}

\author{Sang Won Lee}
\email{sangwonlee@vt.edu}
\orcid{0000-0002-1026-315X}
\affiliation{%
  \institution{Virginia Tech}
  \city{Blacksburg}
  \state{Virginia}
  \country{USA}}

\author{Lifu Huang}
\email{lfuhuang@ucdavis.edu}
\orcid{0000-0002-2743-7718}
\affiliation{%
  \institution{University of California, Davis}
  \city{Davis}
  \state{California}
  \country{USA}}

\author{Pamela Wisniewski}
\email{pamwis@stirlab.org}
\orcid{0000-0002-6223-1029}
\affiliation{%
  \institution{International Computer Science Institute}
  \city{Berkeley}
  \state{California}
  \country{USA}}

\author{Jin-Hee Cho}
\email{jicho@vt.edu}
\orcid{0000-0002-5908-4662}
\affiliation{%
  \institution{Virginia Tech}
  \city{Falls Church}
  \state{Virginia}
  \country{USA}}

\renewcommand{\shortauthors}{An et al.}

\begin{abstract}
Cybergrooming exploits minors through online trust-building, yet research remains fragmented, limiting holistic prevention. Social sciences focused on behavioral insights, while computational methods emphasized detection, but their integration remains insufficient. This review systematically synthesized both fields using the PRISMA framework to enhance clarity, reproducibility, and cross-disciplinary collaboration. Findings showed that qualitative methods offered deep insights but are resource-intensive, machine learning models depend on data quality, and standard metrics struggle with imbalance and cultural nuances. By bridging these gaps, this review advanced interdisciplinary cybergrooming research, guiding future efforts toward more effective prevention and detection strategies.
\end{abstract}




\maketitle

\section{Introduction} \label{sec:introduction}

\new{Cybergrooming is a staged, longitudinal process in which perpetrators exploit online communication to build trust with minors and progressively manipulate them toward sexualized interactions~\cite{mladenovic2021cyber}. It unfolds through identifiable phases, relationship formation, emotional manipulation, boundary testing, and escalation~\cite{oconnell2003typology}, making early behaviors difficult to distinguish from benign interaction without contextual and temporal analysis. Cybergrooming has severe implications for youth, including long-term psychological distress, impaired social trust, and increased risk of offline exploitation. As youth engagement with online platforms continues to grow, these harms have elevated cybergrooming to a central concern in child protection and law enforcement~\cite{ringenberg2022prepost, whittle2013character}.  Despite growing attention, cybergrooming research remains fragmented. Social science studies offer rich accounts of grooming mechanisms but lack computational scalability, while computational approaches prioritize automated detection yet often reduce grooming to static classification tasks, overlooking its staged and adaptive dynamics.  This divide is reflected in existing reviews (Table~\ref{tab:sota-review-papers-analysis}): social science surveys are largely technically agnostic~\cite{ringenberg2022prepost, whittle2013character, schittenhelm2024cybergrooming}, whereas computational surveys emphasize detection pipelines while remaining context-blind to psychological mechanisms~\cite{Ngejane_etal_2018, borj2023chatlog}. Many reviews further subsume cybergrooming under broader online harms~\cite{Razi_etal_2021}, diluting its distinctive process-driven nature.}

\new{To address these gaps, this paper presents a focused interdisciplinary review that integrates social-behavioral insights with computational approaches. By aligning grooming mechanisms with data representations, evaluation practices, and system design, we establish a unified foundation for next-generation detection, prevention, and intervention systems.}

\new{The central takeaway of this review is that social science research conceptualizes cybergrooming as a broader sociotechnical process, encompassing victim risk factors, staged manipulation, and prevention strategies, whereas computational research has largely narrowed the problem space to predator detection as a classification task. Bridging this gap requires computational researchers to expand toward victim-centered risk modeling, stage-aware interventions, and educational prevention systems, rather than treating detection alone as a sufficient solution.} %\jhc{not shown in the response file}

This review makes the following {\bf key contributions}:
\begin{enumerate}
    \item {\bf Interdisciplinary Analysis:} We systematically analyze cybergrooming research across both social and computational sciences, highlighting strengths, limitations, insights, and lessons learned to guide future interdisciplinary research and collaboration.
    \item {\bf Methodological Rigor:} Using the Preferred Reporting Items for Systematic Reviews and Meta-Analyses (PRISMA) methodology~\cite{page2021prisma}, we enhance clarity, transparency, and reproducibility, ensuring a structured analysis of cybergrooming methods, datasets, evaluation metrics, challenges, and future research needs.
    \item {\bf Research Gaps:} Our exclusive focus on cybergrooming enables the identification of distinct research gaps, advocating for stronger interdisciplinary collaboration to enhance understanding and mitigation strategies.
    \item {\bf Multidisciplinary Integration:} We emphasize multidisciplinary integration, demonstrating how insights from social sciences can refine computational models, improve detection accuracy, and support culturally adaptive, real-world solutions for law enforcement, policymakers, and online safety initiatives.
\end{enumerate}

We aimed to answer the following {\bf research questions}:
\begin{enumerate}
\setlength{\itemindent}{2em}
    \item[\bf RQ1] \textit{What are the key approaches used to study cybergrooming in social and computational sciences?}
    \item[\bf RQ2] \textit{What are the primary contributions of existing cybergrooming research in both fields?}
    \item[\bf RQ3] \textit{What limitations exist in the methodologies used across different disciplines?}
    \item[\bf RQ4] \textit{How do social and computational studies on cybergrooming differ, and what interdisciplinary insights can enhance mitigation of cybergrooming risks?}
    \item[\bf RQ5] \textit{What gaps and challenges remain in cybergrooming research, and what future directions are needed?}
\end{enumerate}

We will discuss the answers to these research questions in Section~\ref{sec:discussions}.

\begin{figure}[t]
\centering
\small
\begin{tikzpicture}[
    node distance=1.4cm and 2.2cm,
    >=Stealth,
    box/.style={
        rounded corners=4pt,
        draw=black!70,
        align=center,
        minimum width=4.6cm,
        minimum height=1.2cm,
        font=\small
    },
    foundation/.style={box, fill=blue!15},
    social/.style={box, fill=purple!15},
    computational/.style={box, fill=red!15},
    integration/.style={box, fill=green!15}
]

% Bottom Layer
\node[foundation] (foundation)
{ \textbf{Sections 1--2: Foundations}\\
Phenomenon Definition\\
Systematic Review Methodology};

% Middle Layer
\node[social, above left=0.7cm of foundation] (social)
{ \textbf{Section 3: Social Science}\\
Stages \& Vulnerability\\
Context \& Resistance };

\node[computational, above right=0.7cm of foundation] (comp)
{ \textbf{Section 4: Computational Science}\\
NLP \& ML Detection\\
Behavioral Modeling};

% Top Layer
\node[integration, above=1.5cm of foundation] (integration)
{ \textbf{Sections 5--6: Integration \& Synthesis}\\
Adaptive Socio-Technical Framework};

% Arrows from foundation (orthogonal style)
\draw[->, thick]
    (foundation.west) -|
    node[above, font=\scriptsize, xshift=10mm]{Conceptual Framing} 
    (social.south);

\draw[->, thick]
    (foundation.east) -|
    node[above, font=\scriptsize, xshift=-13mm]{Methodological Structure} 
    (comp.south);

% Bidirectional arrow between social & computational
\draw[<->, thick]
    (social.east) -- 
    node[below, font=\scriptsize]{Theory $\leftrightarrow$ Algorithm Design}
    (comp.west);

% Arrows to integration (orthogonal style)
\draw[->, thick]
    (social.north) |-
    node[above, font=\scriptsize]{Risk Models \& Stage Theory}
    (integration.west);

\draw[->, thick]
    (comp.north) |-
    node[above, font=\scriptsize]{Detection Signals \& Scalability}
    (integration.east);

\end{tikzpicture}

\caption{\new{Socio-technical structure of the paper. Sections~1--2 establish conceptual and methodological foundations, Section~3 develops social-scientific models, Section~4 advances computational detection approaches, and Sections~5--6 synthesize these domains into an adaptive interdisciplinary framework.}}
\label{fig:paper_structure}
\end{figure}

\new{Fig.~\ref{fig:paper_structure} presents the paper as a layered socio-technical framework. Sections~1--2 establish conceptual framing and methodological structure, forming the foundation for domain-specific analysis. Section~3 articulates social-scientific insights into grooming stages, vulnerability, and resistance, while Section~4 operationalizes these insights through computational modeling, NLP, and scalable detection systems. The bidirectional exchange between these domains highlights the mutual reinforcement of theory and algorithm design. Sections~5--6 integrate both perspectives, consolidating risk models and detection mechanisms into an adaptive interdisciplinary framework aimed at advancing coordinated cybergrooming prevention.
}

\section{Review Methodology} \label{sec:review-methodology}

\subsection{The Preferred
Reporting Items for Systematic reviews
and Meta-Analyses (PRISMA)}

\new{We conducted the literature review following the Preferred Reporting Items for Systematic Reviews and Meta-Analyses (PRISMA) 2020 guidelines~\cite{page2021prisma} to ensure transparency, methodological rigor, and reproducibility. PRISMA provides a standardized structure for reporting literature search, screening, and eligibility decisions for interdisciplinary cybergrooming research spanning qualitative social science studies and computational approaches.}

\subsection{Search Strategy and Information Sources} \label{subsec:info-sources}

\new{We conducted the literature search in two stages using identical databases, search queries, and inclusion criteria to ensure coverage and temporal relevance.}

\noindent\new{\textbf{Initial search.}  The initial search covered studies published up to September 30, 2024. Social science databases included PsycINFO, PubMed, SocINDEX, ERIC, and Web of Science, while computational science sources included the ACM Digital Library, IEEE Xplore, Computers \& Applied Sciences Complete, and ScienceDirect. Titles and abstracts were queried using:
\textit{(``online grooming'' OR ``cybergrooming'' OR ``online sexual grooming'' OR ``internet grooming'' OR ``cyber grooming'')}.} \new{This search returned 770 candidate records across social and computational science venues (360 social science; 410 computational science).}

\noindent\new{\textbf{Updated search.}  To incorporate newly published work, we conducted an updated search covering October 1, 2024, through December 1, 2025, yielding 182 additional records (92 social science; 90 computational science).  Because some databases do not support month-level filtering, all records indexed during 2024--2025 were retrieved, and previously reviewed studies were removed via cross-stage deduplication. Identical screening and eligibility criteria were applied across both search stages to ensure consistency.}

To ensure completeness, we conducted a supplementary search to capture any relevant publications that may have been initially overlooked. This included a citation analysis of key studies, identifying recent works citing foundational sources, and manually reviewing reference lists. Additionally, we searched Google Scholar and arXiv for non-peer-reviewed studies, preprints, and emerging research that might not yet be indexed in traditional databases. This supplementary search yielded 25 additional relevant papers, further strengthening the comprehensiveness of our review.

\begin{wrapfigure}{r}{0.55\textwidth}
    \vspace{-10mm}
    \centering
    \resizebox{\linewidth}{!}{ % Dynamically scales the TikZ picture to the wrapfigure width
    \begin{tikzpicture}[>=latex, font={\sf \normalsize}]

    \tikzstyle{bluerect} = [rectangle, rounded corners, minimum width=1.2cm, minimum height=0.6cm, text centered, draw=black, fill=cyan!60!gray!45!white, rotate=90, font=\sffamily]

    \tikzstyle{roundedrect} = [rectangle, rounded corners, minimum width=8cm, minimum height=0.8cm, text centered, draw=black, font=\sffamily]

    \tikzstyle{textrect} = [rectangle, minimum width=4.8cm, text width=4.8cm, minimum height=0.8cm, draw=black, font={\sffamily \scriptsize}]

    \tikzstyle{textrects} = [rectangle, minimum width=5.6cm, text width=5.6cm, minimum height=0.8cm, draw=black, font={\sffamily \scriptsize}]

    \node (r1blue) at (-5.5cm, 6.0cm) [draw, bluerect, minimum width=1.5cm, font=\small]{Identification};

    \node (r1left) at (-2.5cm, 6.0cm) [draw, textrect, minimum height=1.6cm, font=\small]
      {Records identified from:
         \begin{itemize}
         \item Databases ($n=770$, \nd{$n=182$})
         \item Additional Search ($n=25$)
         \end{itemize}   
      };

    \node (r1right) at (3.5cm, 6.0cm) [draw, textrects, minimum height=1.5cm, font=\small]
      {Records removed before screening: 
        \begin{itemize}
        \item Duplicates Zotero ($n=192$, \nd{$n=22$}) 
        \item Duplicates Manual ($n=20$, \nd{$n=36$})
        \item Retracted ($n=1$)
        \end{itemize} 
      };

    \node (r2blue) at (-5.5cm, 2.5cm) [draw, bluerect, minimum width=4cm, font=\small]
      {Screening};

    \node (r2left) at (-2.5cm, 4.0cm) [draw, textrect, minimum height=1.2cm, font=\small]
      {Publications screened \\ ($n=582$, \nd{$n=131$})};

    \node (r2right) at (3.5cm, 4.0cm) [draw, textrects, minimum height=1.2cm, font=\small]
      {Publications excluded \\ ($n=408$, \nd{$n=81$})};

    \node (r3left) at (-2.5cm, 2.5cm) [draw, textrect, minimum height=1.2cm,font=\small]
      {Publications sought for retrieval \\ ($n=174$ , \nd{$n=43$})};

    \node (r3right) at (3.5cm, 2.5cm) [draw, textrects, minimum height=1.2cm,font=\small]
      {Not retrieved \\ ($n=4$, \nd{$n=1$})};

    \node (r4left) at (-2.5cm, 1cm) [draw, textrect, minimum height=1.2cm,font=\small]
      {Publications assessed for eligibility \\ ($n=170$, \nd{$n=42$})};

    \node (r4right) at (3.5cm, -0.4cm) [draw, textrects, minimum height=1.5cm,font=\small]
      {Publications Excluded ($n=91$, \nd{$n=29$}): 
        \begin{itemize}
        \item Wrong language ($n=10$, \nd{$n=6$})
        \item Review article ($n=17$, \nd{$n=4$})
        \item Incorrect format ($n=4$, \nd{$n=2$})
        \item Cybergrooming as a minor focus ($n=49$, \nd{$n=10$})
        \item Offline grooming ($n=10$, \nd{$n=1$})
        \item Not focus on either victim or offender ($n=1$, \nd{$n=6$})
        \end{itemize}     
      };

    \node (r5blue) at (-5.5cm, -1.2cm) [draw, bluerect, minimum width=1.6cm,font=\small]
      {Included};

    \node (r5left) at (-2.5cm, -1.2cm) [draw, textrect, minimum height=1.6cm,font=\small]
      {Publications included in review \\
       ($n=79$, \nd{$n=13$}) \\ };

    % Draw arrows between nodes:
    \draw[thick, ->] (r1left) -- (r1right);
    \draw[thick, ->] (r1left) -- (r2left);
    \draw[thick, ->] (r2left) -- (r2right);
    \draw[thick, ->] (r2left) -- (r3left);
    \draw[thick, ->] (r3left) -- (r3right);
    \draw[thick, ->] (r3left) -- (r4left);
    \draw[thick, ->] (r4left.east) -- (r4right.west |- r4left.east);
    \draw[thick, ->] (r4left) -- (r5left);

    \end{tikzpicture}
    } % End resizebox
    \caption{\new{PRISMA flow diagram of the study selection process (black: initial search; green: updated search).}}
    \label{fig:PRISMA}
    \vspace{-3mm}
\end{wrapfigure}

\subsection{Review Criteria and PRISMA Result} \label{subsec:review-criteria}

\new{\textbf{Inclusion and exclusion criteria.} Studies were included if they: (1) focused on cybergrooming as a primary analytical topic; (2) involved minors as the target population; (3) were peer-reviewed and written in English; and (4) contributed empirical or computational insights. We excluded studies that addressed only offline grooming, non-sexual uses of the term grooming, review-only papers, or works lacking analysis of victim or offender behavior.}

\noindent\new{\textbf{Screening process and final corpus.}  From the initial search, 795 records were reduced to 582 after deduplication. Title and abstract screening excluded 408 records due to irrelevance or ambiguous use of the term \textit{grooming}.}

\new{The updated search underwent the same screening procedure. After deduplication, 124 records remained for title screening, 43 proceeded to full-text review, and 13 studies met all inclusion criteria (5 computational; 8 social science).  All newly identified studies were integrated into the final corpus. The complete screening workflow and the distinction between the two search stages are shown in Fig.~\ref{fig:PRISMA}.}

\begin{wrapfigure}{r}{0.55\textwidth}
    \vspace{-5mm}
    \centering
    \begin{tikzpicture}
    \pgfplotsset{compat=1.16}
    \begin{axis}[
        ybar=0.5pt,
        ymin=0, ymax=14,
        height=6cm,
        width=\linewidth,
        bar width=6pt, 
        ylabel={Number of Works},
        xlabel={Publication Year},
        ylabel style={font=\footnotesize},
        xlabel style={font=\footnotesize},
        tick label style={font=\scriptsize},
        xtick={2009,2010,2011,2012,2013,2014,2015,2016,2017,2018},
        % Angled labels to prevent overlapping in a narrow column
        xticklabel style={rotate=45, anchor=north east, inner sep=2pt, font=\scriptsize}, 
        xticklabels={2007-08, 2009-10, 2011-12, 2013-14, 2015-16, 2017-18, 2019-20, 2021-22, 2023-24, 2025*},
        nodes near coords,
        nodes near coords style={font=\tiny}, % Scaled down data labels
        nodes near coords align={vertical},
        % Legend moved to the top to save horizontal space inside the graph
       % legend style={font=\scriptsize, at={(0.5,1.02)}, anchor=south, legend columns=2, draw=none},
       legend style={
    font=\scriptsize,
    at={(0.45,0.9)},
    anchor=north east,
    legend columns=1,
    draw=white,
    fill=white,
    fill opacity=0.8,
    text opacity=1
},
        enlarge x limits=0.08 % Adds a tiny bit of padding on the left and right
    ]
    
    \addplot[fill=blue!30, draw=blue, nodes near coords style={color=blue}] 
        coordinates {(2009,2)(2010,1)(2011,1)(2012,2)(2013,5)(2014,5)(2015,6)(2016,9)(2017,10)(2018,8)};
        
    \addplot[fill=red!30, draw=red, nodes near coords style={color=red}] 
        coordinates {(2009,0)(2010,1)(2011,3)(2012,3)(2013,4)(2014,1)(2015,6)(2016,7)(2017,11)(2018,5)};
        
    \legend{Social Sciences, Computational Sciences}
    \end{axis}
    \end{tikzpicture}
    \vspace{-4mm}
    \caption{\new{Trends in social and computational science articles on cybergrooming (2007–2025; 2025* includes publications through Dec 1 only).}}
    \label{fig:Literatures_numbers}
    \vspace{-4mm}
\end{wrapfigure}

\textbf{Research trends.} Fig.~\ref{fig:Literatures_numbers} illustrates the growth of cybergrooming research over time. Both social and computational sciences show increasing contributions, with a sharp rise in computational publications from 2015 onward. \new{Social science research on cybergrooming has grown steadily, driven by rising concern for youth online safety, increased data availability, and heightened attention to child protection and digital ethics. This growth reflects a shift toward systematic empirical methods, including longitudinal, validated, and cross-national studies of psychological, social, and contextual risk factors. In parallel, computational research has accelerated with advances in AI, together representing complementary responses to evolving child-safety challenges.}

Beyond the overall growth, the rise in social science publications after 2015 reflects the availability of more robust empirical data on cybergrooming. \new{Population-based and cross-national studies documenting prevalence~\cite{Finkelhor22-online-sexual-offense, schittenhelm2024cybergrooming}, adolescents’ routine online exposure~\cite{wachs2020routine}, and risks linked to intensive digital engagement~\cite{almeida2024online} enable rigorous behavioral and psychosocial analyses, establishing cybergrooming as a central issue in child protection and developmental research.}

\begin{table}[t]
\small 
\centering
\caption{Summary of Related Review Papers on Cybergrooming Research} \label{tab:sota-review-papers-analysis}
\vspace{-3mm}
\begin{tabular}{p{3cm} p{4cm} p{3cm} p{4cm}}
\hline
\multicolumn{1}{c}{\bf Review Paper} & \multicolumn{1}{c}{\bf Key Contribution} & \multicolumn{1}{c}{\bf Limitations} & \multicolumn{1}{c}{\bf Differences with Our Review} \\ \hline
\multicolumn{4}{c}{\textbf{\gray Social Sciences}}
\\ \hline
\citet{ringenberg2022prepost} & Reviewed grooming techniques before and after the advent of the Internet, highlighting similarities in strategies. & Focused only on groomers' strategies without addressing detection methods. & Our review explores potential solutions for online grooming, including detection strategies, offering a more solution-oriented perspective. \\ \hline
\citet{whittle2013character} & Analyzed online grooming characteristics and younger generations' behavior and consistency in the grooming techniques. & Descriptive in nature, without proposing solutions to counter cybergrooming. & Our review provides actionable insights and strategies for addressing cybergrooming, filling the gap left by descriptive studies. \\ \hline
\citet{schittenhelm2024cybergrooming} & Conducted a systematic review of cybergrooming victimization, examining prevalence, risk factors, and outcomes. & Relies on cross-sectional, self-reported data, potentially overlooking minority groups and limiting causal insights. & Our review takes an interdisciplinary approach, integrating computational methods with social science insights, broadening the scope beyond victimization. \\ \hline
\multicolumn{4}{c}{\textbf{\gray Computational Sciences}}
\\ \hline
\citet{Ngejane_etal_2018} & Reviewed machine learning techniques in cybersecurity to combat online grooming, emphasizing supervised learning. & Focusing exclusively on machine learning without considering the broader social context. & We take an interdisciplinary approach, combining computational and social sciences for a holistic view of cybergrooming. \\ \hline
\citet{Razi_etal_2021} & Examined computational methodologies for detecting online sexual risks, emphasizing dataset quality and real-world applicability. & Broad scope, covering various online risks (grooming, trafficking, harassment). & Our review specifically targets cybergrooming, providing a focused analysis relevant for researchers interested in this area. \\ \hline
\citet{borj2023chatlog} & Analyzed detection methods for online grooming using chat log data, focusing on detection algorithms and datasets. & Primarily focuses on psychological theories and grooming detection. & Our review synthesizes insights across social and computational sciences, offering a multidisciplinary understanding of cybergrooming. \\ \hline
\citet{mladenovic2021cyber} & Surveyed definitions and linguistic features in cyber-aggression, covering cybergrooming as part of broader cyber abuse. & Broad coverage of cyber abuse topics, with cybergrooming as a minor focus. & Our review focuses on cybergrooming, providing an in-depth resource for studying online sexual exploitation on social media. \\ \hline
\end{tabular}
\vspace{-5mm}
\end{table}

\subsection{Other Related Review Papers} \label{subsec:other-cybergrooming-review}

\new{\textbf{Social science oriented reviews.}
Existing social science reviews examine cybergrooming primarily as a behavioral and psychosocial process, emphasizing grooming strategies, victim vulnerability, and developmental dynamics~\cite{ringenberg2022prepost, whittle2013character, schittenhelm2024cybergrooming}. While theoretically and empirically grounded, these reviews remain largely explanatory, with limited engagement in computational detection or intervention mechanisms.}

\new{\textbf{Computationally oriented reviews.}
Computational surveys focus on machine learning–based detection methods, datasets, and algorithmic pipelines for online sexual risk analysis~\cite{Ngejane_etal_2018, borj2023chatlog}, often treating cybergrooming as one of several online risks~\cite{mladenovic2021cyber}, which constrains grooming-specific synthesis.}

\new{\textbf{Cross-disciplinary limitations.}
Overall, prior reviews reflect a persistent disciplinary divide: social science work advances behavioral insight, while computational surveys emphasize technical detection with limited theoretical grounding. Reviews spanning multiple online harms further dilute cybergrooming-specific analysis, motivating the need for an integrative synthesis bridging behavioral theory and computational approaches (Table~\ref{tab:sota-review-papers-analysis}).}

\section{Cybergrooming Research in Social Sciences} \label{sec:cybergrooming-SC}

\new{This section organizes prior studies by research objectives, methods, and evaluation practices, noting relevant theoretical foundations where they inform design and interpretation.}

\subsection{Research Objectives in Social Science Cybergrooming Research} \label{subsec:ss-research-objectives}

\new{Social Science approaches to cybergrooming research center on four primary objectives:}

\cbox{yellow!90!black} \textbf{Risk Analysis.}~\cite{gamez2023stability, whittle2014their, sani_social_2021, kamar2024relevance, Chiu2022onlinegrooming, almeida2024online, Finkelhor22-online-sexual-offense, HERNANDEZ2021106569, wachs2020routine, Wachs2012, Weingraber2020,  wachs2016cross, gamez-guadix_construction_2018, schoeps2020risk, gamez2018persuasion, villacampa2017online, lorenzo2016understanding, quayle2014rapid, thomas2023offenders, almeida2025prevalence, melo2025grooming, reneses2024he, aktu2024self, lorenzo5280035not, resett2025prediction} Social science research extensively examined the underlying risk factors contributing to cybergrooming victimization. \new{These studies examine associations between cybergrooming, sexting, sextortion, and parental control measures, highlighting structural vulnerabilities in adolescent online environments~\cite{almeida2024online}.} Other research identified demographic and behavioral factors influencing victim susceptibility (e.g., age, gender, and routine online behaviors)~\cite{Finkelhor22-online-sexual-offense, wachs2020routine}. Emotional consequences (e.g., shame, guilt, psychological distress) have also been examined to understand coping responses~\cite{gamez2023stability, HERNANDEZ2021106569}. These findings inform policies to reduce risk exposure and strengthen online safety.

\cbox{teal!20} \textbf{Offender Profiling.}~\cite{Anggraeny2023, Fernandes_et_al_2023, Lorenzo-Dus_Izura_2017, gamez2021unraveling, van2016behavioural, kloess2019case, villacampa2017online, de_santisteban_progression_2018, broome_psycho-linguistic_2020, soldino_criminological_2024, broome_investigation_2024, lorenzo2016understanding, quayle2014rapid, thomas2023offenders, gamez2018persuasion, shannon2008sweden, lorenzo2020modus, seymour2021discursive} This topic deals with cybergrooming patterns and offender behaviors, categorizing stages and tactics used by predators. These studies have examined linguistic strategies, manipulation techniques, and grooming progression to build offender typologies \cite{Anggraeny2023, Fernandes_et_al_2023}. Specific tactics, such as strategic praise and emotional manipulation, reveal how predators gain victims’ trust before exploitation.% \cite{Lorenzo-Dus_Izura_2017, gamez2021unraveling}. %This knowledge aids early detection and informs law enforcement efforts.

\cbox{blue!60} \textbf{Prevention and Awareness.}~\cite{Calvete_Orue_GamezGuadi_2022, Dorasamy_etal_2021, carmo2023knowledge, choo_responding_2009, kamar2022parental, dolev-cohen_parental_2024, craven_current_2007, christiansen2024hidden, khotimah2024child, lundrigan2025relationship} Raising awareness and educating victims, parents, and educators about cybergrooming is a key focus of social science research. Studies evaluate educational programs that enhance online safety knowledge among minors and caregivers \cite{Calvete_Orue_GamezGuadi_2022}. Other research highlights parental supervision strategies in mitigating grooming risks \cite{Dorasamy_etal_2021, kamar2022parental, christiansen2024hidden, khotimah2024child, lundrigan2025relationship}. Additionally, scholars assess awareness campaigns and school-based digital literacy initiatives in reducing victimization rates \cite{craven_current_2007}.

\cbox{red!70} \textbf{Behavioral Stages.}~\cite{black2015linguistic, Anggraeny2023, grant_assuming_2016, gamez2018persuasion, evans2025corpus} Understanding the sequential phases of cybergrooming is crucial for intervention and detection. Research categorizes grooming into stages such as access, relationship-building, manipulation, exploitation, and concealment. Linguistic and behavioral analyses compare online grooming strategies with offline tactics, identifying distinct features of digital interactions. These findings help refine detection models and strengthen response strategies for law enforcement and child protection agencies.

\begin{table}[htbp] % Adjusted float positioning
\small
    \caption{Research Objectives of the Reviewed Social Science Research Works on Cybergrooming}
    \label{tab:research-objects-ss}
    \vspace{-3mm}
    \begin{tabular}{p{4cm} P{2.5cm} P{2.5cm}} % Adjusted column widths
    \hline
        \textbf{Research Objectives} & \textbf{Percentage} & \textbf{Number of Articles} \\
    \hline
        \raisebox{-0.05cm}{\cbox{yellow!90!black}} Risk Analysis & \new{43.11\%} & \new{25} \\
    \hline
        \raisebox{-0.05cm}{\cbox{teal!20}} Offender Profiling & \new{31.03\%} & \new{18} \\
    \hline
        \raisebox{-0.05cm}{\cbox{blue!60}} Prevention and Awareness & \new{17.24\%} & \new{10} \\
    \hline
        \raisebox{-0.05cm}{\cbox{red!70}} Behavioral Stages & \new{8.62\%} & \new{5} \\
    \hline
    \end{tabular}
    \vspace{1mm}

    \small{(Note: An article may appear multiple times if it addresses multiple objectives, each counted individually.)}
    \vspace{-3mm}
\end{table}

Table~\ref{tab:research-objects-ss} shows that most research in social sciences on cybergrooming focuses on \textit{risk analysis} (\new{43.11\%}) and \textit{offender profiling} (\new{31.03\%}). \textit{Prevention and awareness} (\new{17.24\%}) and \textit{behavioral stages} (\new{8.62\%}) are less frequently addressed, indicating a primary focus on understanding risks and offender profiles over awareness initiatives and intervention planning.

A limitation of existing research is the relatively lower emphasis on \textit{prevention and awareness} and \textit{behavioral stages}, critical for proactive intervention. This gap highlights the need to bridge theoretical findings with practical applications, such as integrating awareness programs into law enforcement training and developing real-time intervention tools.

To address these gaps, social science research on cybergrooming could benefit from techniques developed in computational sciences, where detection-focused approaches dominate. Computational methods, including machine learning models and natural language processing, can enhance the identification of grooming stages and provide scalable tools for raising awareness. By integrating these techniques, social science research could strengthen preventive strategies and foster a more proactive response to cybergrooming.
\subsection{Risk Factors of Victims to Cybergrooming Identified in Social Science Cybergrooming Research} \label{subsec:ss-risk-factors}

\begin{wrapfigure}{r}{0.7\textwidth}
\vspace{-8pt}
\centering
\resizebox{\linewidth}{!}{
\begin{tikzpicture}[
    font=\scriptsize,
    >=Stealth,
    node distance=0.9cm,
    every node/.style={align=center}
]

% =========================
% Top row (light → darker blue)
% =========================

\node[draw, rounded corners=3pt, fill=blue!15,
      text width=2cm, minimum height=0.6cm] (beh)
{Behavioral \& Social Factors};

\node[draw, rounded corners=3pt, fill=blue!25,
      right=0.2cm of beh,
      text width=2cm, minimum height=0.6cm] (pers)
{Personal \& Demographic Factors};

\node[draw, rounded corners=3pt, fill=blue!40,
      right=0.2cm of pers,
      text width=2cm, minimum height=0.6cm] (psych)
{Psychological \& Emotional Factors};

\node[draw, rounded corners=3pt, fill=blue!60, text=white,
      right=0.2cm of psych,
      text width=2cm, minimum height=0.6cm] (tech)
{Technological Factors};

% =========================
% Middle row (lightest → darkest red)
% =========================

\node[draw, fill=orange!15!,
      below=0.7cm of beh,
      text width=2.5cm, minimum height=0.7cm, xshift = 7mm] (exp)
{Increased Online Exposure};

\node[draw, fill=orange!40,
      right=0.5cm of exp, 
      text width=2.5cm, minimum height=0.7cm] (par)
{Reduced Parental Supervision};

\node[draw, fill=orange!85, 
      right=0.5cm of par,
      text width=2.5cm, minimum height=0.7cm] (vul)
{Higher Psychological Vulnerability};

% =========================
% Bottom box (black)
% =========================

\node[draw, rounded corners=4pt,
      fill=black, text=white,
      below=0.5cm of par,
      minimum width=3.4cm, minimum height=0.7cm] (risk)
{\bf Increased Cybergrooming Risk};

% =========================
% Arrows
% =========================

\draw[->, thick, blue!15!black] (beh.south) -- (exp.north);
\draw[->, thick, dashed, blue!25!black] (pers.south) -- (par.north);
\draw[->, thick, blue!40!black] (psych.south) -- (vul.north);

% Two blue arrows from Technological Factors
\draw[->, thick, blue] (tech.south) -- (vul.north);
\draw[->, thick, blue, dashed] (tech.south) -- (exp.north);
\draw[->, line width=1.6pt, orange!15!black] (exp.south) -- (risk.west);
\draw[->, line width=1.6pt, orange!40!black] (par.south) -- (risk.north);
\draw[->, line width=1.6pt, orange!85!black] (vul.south) -- (risk.east);

\end{tikzpicture}
}
\vspace{-7mm}
\caption{Conceptual model of cybergrooming risk factors: Interaction between individual, environmental, and technological influences.} \label{fig:risk-factors-model}
\vspace{-7mm}
\end{wrapfigure}
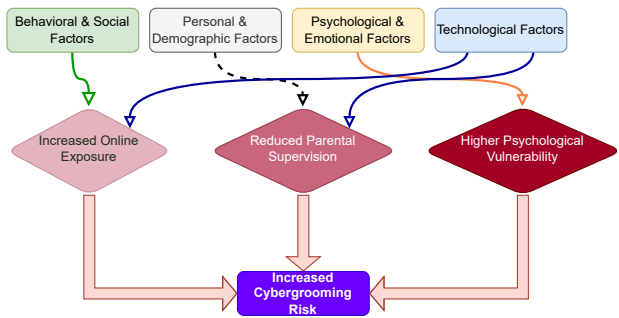

Research on cybergrooming has identified multiple victim characteristics associated with heightened risk. These factors can be broadly grouped into \textbf{behavioral and social}, \textbf{personal and demographic}, \textbf{psychological and emotional}, and \textbf{technological} domains, each contributing to adolescent vulnerability. Fig.~\ref{fig:risk-factors-model} presents a conceptual framework illustrating the interactions among these categories and showing how individual, environmental, and technological influences jointly shape susceptibility to cybergrooming.

\begin{itemize}
    \item \textbf{Behavioral and Social Factors:} Engaging in \textit{Online Disclosure}~\cite{wachs2020routine, whittle2014their}, such as sharing private information and maintaining open profiles, increases cybergrooming risk. Adolescents under restrictive \textit{Parental Mediation}~\cite{wachs2020routine} tend to disclose more personal information online compared to those under instructive mediation, thereby heightening their vulnerability. \new{In addition, insufficient protective social ties with friends and family further increase vulnerability to grooming risks~\cite{lundrigan2025relationship}.} Additionally, individuals with a \textit{Past Experience} of cyberbullying~\cite{Wachs2012} or a \textit{Willingness to Meet Strangers}~\cite{Wachs2012, whittle2014their} face significantly higher risks. Those who \textit{Spend extensive time online}~\cite{whittle2014their} also show an increased likelihood of encountering groomers.

    \item \textbf{Personal and Demographic Factors:} \textit{Gender}~\cite{Finkelhor22-online-sexual-offense, Wachs2012} and \textit{Age}~\cite{gamez2023stability, almeida2025prevalence, resett2025prediction} play a crucial role, with older adolescents generally facing greater risks. \new{While much of the literature reports higher victimization among females, emerging regional findings suggest that older adolescents and males may experience higher frequencies of predatory contact in certain contexts~\cite{resett2025prediction}.}
    Additional demographic factors include \textit{Foreign birth}~\cite{gamez2023stability} and \textit{Sexual minority identification}~\cite{gamez2023stability}, both of which have been associated with higher vulnerability. Family circumstances, such as \textit{Parents with lower education levels}~\cite{gamez2023stability}, \textit{Living with separated or divorced parents}~\cite{gamez2023stability, whittle2014their}, \textit{Difficult family relations}~\cite{whittle2014their}, \new{or \textit{Economic hardship}~\cite{melo2025grooming}}, further elevate susceptibility.
    A \textit{Lack of education about online safety}~\cite{whittle2014their}, whether at home or in school, also contributes to increased exposure to online grooming risks.

    \item \textbf{Psychological and Emotional Factors:} Adolescents at higher risk often experience elevated \textit{anxiety}, \textit{depression}~\cite{gamez2023stability, gamez2021unraveling}, \textit{shame}~\cite{gamez2023stability}, and \textit{guilt}~\cite{gamez2023stability}.
    \new{Recent evidence further links grooming vulnerability to depressive symptoms, anxiety, suicidality, and reduced well-being among adolescents with abuse histories~\cite{melo2025grooming}.} These psychological factors make adolescents more susceptible to manipulative grooming tactics. \textit{Low self-esteem}~\cite{wachs2016cross, whittle2014their, aktu2024self} and \textit{loneliness}~\cite{whittle2014their} further increase vulnerability.
    \textit{Disinhibition}~\cite{HERNANDEZ2021106569}, reducing self-restraint in online interactions, heightens susceptibility. Specific personality traits such as \textit{extraversion} and \textit{lack of empathy}~\cite{HERNANDEZ2021106569} in boys and \textit{narcissism}~\cite{HERNANDEZ2021106569} in girls have also been linked to greater grooming risk. \new{Gender differences in risk perception further influence vulnerability, as boys are more likely to underestimate online risks, while offenders strategically use flattery to construct false intimacy~\cite{reneses2024he}.}
    
    \item \textbf{Technological Factors:} \new{Adolescents’ access to personal smartphones increases cybergrooming risk by enabling continuous, private, and weakly supervised communication~\cite{wachs2020routine, whittle2014their, lundrigan2025relationship}. In contemporary digital settings, risk is shaped less by physical access location and more by app-based features that support secrecy and sustained interaction, including encryption, disappearing messages, and multi-platform communication that can bypass parental oversight~\cite{lundrigan2025relationship, aktu2024self}. Recent studies further link grooming susceptibility to problematic Internet usage and digitally mediated intimacy, which can amplify psychological and familial vulnerabilities~\cite{aktu2024self}.}

\end{itemize}
The identified characteristics provide valuable insights into the risk factors associated with cybergrooming, highlighting how behavioral patterns, social environments, demographic variables, psychological traits, and technological access collectively shape adolescent vulnerability. Frequent online disclosure and a lack of parental guidance are significant behavioral risks, while emotional instability, such as low self-esteem and loneliness, further exacerbates susceptibility. Additionally, access to private internet spaces increases the opportunity for prolonged grooming attempts. 

Fig.~\ref{fig:risk-factors-model} illustrates the conceptual model of cybergrooming risk factors, highlighting how behavioral, demographic, psychological, and technological influences contribute to increased online exposure, reduced parental supervision, and heightened psychological vulnerability. These intermediate factors collectively increase the overall risk of cybergrooming victimization.

However, despite these insights, certain limitations exist. Most available data rely on cross-sectional, self-reported measures, which may introduce biases and limit causal inference. Furthermore, existing research primarily focuses on Western adolescent populations, raising concerns about the generalizability of findings to diverse cultural contexts. Notably, there is a lack of research on how {\em age-specific and cultural variations} influence grooming susceptibility. For example, younger children may be more trusting, whereas older adolescents might engage in riskier online behaviors due to increased autonomy. Similarly, cultural differences in parenting styles and internet use norms may alter vulnerability patterns. Addressing these gaps requires more longitudinal and cross-cultural studies to better understand cybergrooming risks across different social and geographical contexts.

\subsection{Evaluation Methods and Metrics in Social Science Cybergrooming Research} \label{subsec:ss-evaluation-methods}

\new{Across the reviewed literature, methodological choices are often informed by established theoretical perspectives, which guide how grooming behaviors are conceptualized, operationalized, and evaluated. Accordingly, this subsection highlights how such theoretical influences are reflected, where applicable, in evaluation practices across quantitative, qualitative, and mixed-method studies. Consistent patterns emerge, with survey-based quantitative analyses predominating in large-scale studies and qualitative approaches providing contextual insight into grooming behaviors and prevention strategies. Common limitations persist across methods, including reliance on self-reported data and predominantly cross-sectional designs.} \new{Table~\ref{tab:evaluation-metrics-social} in Appendices provides a detailed comparison of these methods.}

\textbf{Quantitative methods}: Surveys and structured questionnaires are frequently used to collect large-scale data on victimization experiences, online behaviors, and demographic influences. Longitudinal and cross-sectional studies track behavioral trends over time, enhancing understanding of risk factors. For instance, \citet{almeida2024online} used surveys to examine online sexual behaviors, while \citet{gamez2023stability} analyzed the link between cybergrooming prevalence and mental health in 746 adolescents. \citet{Finkelhor22-online-sexual-offense} conducted a large-scale survey of 2,639 adolescents to assess online sexual offenses, and \citet{wachs2020routine}, \new{drawing on Routine Activity Theory~\cite{cohen1979social},} collected data from 5,938 teenagers across six countries to study risk factors. \new{Almeida and Pereira~\cite{almeida2025prevalence} adapted and validated the Questionnaire for Online Sexual Solicitation and Interactions with Adults (QOSSIA)~\cite{gamez-guadix_construction_2018}.} These methods enable statistical analyses such as chi-square tests, regression models, and correlation assessments to determine relationships between key variables. \new{More advanced techniques, including structural equation modeling with bootstrapped mediation analysis, have further identified indirect pathways explaining substantial variance in grooming outcomes~\cite{aktu2024self}.} However, their reliance on self-reported data may introduce bias, and imbalanced datasets can affect result accuracy.

\textbf{Qualitative methods}: Document analysis and interviews provide deep insights into the social and psychological aspects of cybergrooming. Document analysis examines legal policies, existing research, and judicial records to identify gaps in regulation and enforcement. For example, \citet{Anggraeny2023} analyzed court decisions, child protection laws, and criminal codes to understand the legal framework surrounding cyber child grooming. Interviews with victims, parents, educators, and law enforcement offer firsthand perspectives on online risks and grooming tactics. \citet{Dorasamy_etal_2021}, \new{drawing on Social Cognitive Theory~\cite{bandura1989human},} interviewed parents to explore their awareness of online grooming risks, while \citet{Fernandes_et_al_2023}, \new{utilizing grounded theory~\cite{glaser2017discovery},} interviewed a law enforcement officer to develop a specialized investigative framework. Thematic analysis~\cite{Dorasamy_etal_2021} is frequently applied to extract recurring themes in interview data, enhancing interpretability. \new{In addition, categorical content analysis supports triangulation across victim, offender, and expert perspectives, strengthening methodological rigor~\cite{reneses2024he}.} \new{Facework-based politeness frameworks have also been applied to code child resistance strategies across conversational turns, enabling fine-grained analysis of interactional dynamics~\cite{lorenzo5280035not}.} While qualitative research provides depth and contextual richness, findings are often difficult to generalize due to smaller sample sizes.

\textbf{Comparing Methods:} Quantitative approaches are most effective for identifying \textit{prevalence rates} and \textit{demographic trends}, while qualitative methods are superior for exploring \textit{psychological motivations}, \textit{perpetrator strategies}, and \textit{legal challenges}. However, an integrated approach combining both methods offers the most robust insights. For example, mixed-methods research can use surveys to establish statistical trends, followed by interviews to explain underlying causes. \citet{Fernandes_et_al_2023} combined literature reviews, thematic analysis, and comparative analysis to construct a comprehensive cybergrooming investigation framework. Similarly, \citet{Anggraeny2023} used deductive reasoning to analyze grooming stages, complementing legal document reviews with offender interviews.

\textbf{Evaluation Metrics:} Statistical tools validate findings and quantify relationships between key variables. Commonly used software includes R~\cite{gamez2023stability, Wachs2012}, SPSS~\cite{HERNANDEZ2021106569, wachs2020routine}, Stata~\cite{Finkelhor22-online-sexual-offense}, and Mplus~\cite{HERNANDEZ2021106569, wachs2020routine}. Chi-square tests, frequency counts, and t-tests assess group differences, while more advanced models rely on Comparative Fit Index (CFI) and Root Mean Square Error of Approximation (RMSEA) to evaluate model fit. \new{Table~\ref{tab:evaluation-metrics-social} in the appendices} details these and additional metrics, such as Cronbach’s alpha for internal consistency and regression coefficients for predictive analysis.

A key takeaway is that \textbf{\em neither method alone is sufficient} to fully understand cybergrooming. While surveys can detect statistical patterns, interviews uncover emotional and behavioral nuances. Therefore, \textbf{\em future research should adopt interdisciplinary, mixed-methods approaches} to integrate large-scale quantitative analyses with rich qualitative insights, improving both the validity and applicability of cybergrooming studies.

\subsection{Evaluation Datasets Used in Social Science Cybergrooming Research} \label{subsec:ss-eval-datasets}

This section describes the datasets commonly used in cybergrooming research within the social sciences:

\ctriangle{red} \textbf{Questionnaire Data}~\cite{almeida2024online, Calvete_Orue_GamezGuadi_2022, Finkelhor22-online-sexual-offense, wachs2020routine, HERNANDEZ2021106569, Wachs2012, Weingraber2020, wachs2016cross, gamez2018persuasion, schoeps2020risk, carmo2023knowledge, villacampa2017online, gamez-guadix_construction_2018, almeida2025prevalence, lundrigan2025relationship, melo2025grooming, aktu2024self, resett2025prediction}: Social science research frequently uses questionnaire data, including longitudinal designs, to examine specific questions and track changes over time. For example, one study surveyed 2,639 respondents to analyze online sexual offenses~\cite{Finkelhor22-online-sexual-offense}, while another used pre-test, 3- and 6-month follow-ups to assess an educational intervention on cybergrooming awareness~\cite{Calvete_Orue_GamezGuadi_2022}. Although such datasets provide large-scale insights, self-reported measures introduce biases such as recall errors and social desirability. To address this, researchers use \textit{data triangulation}, combining surveys with behavioral logs or parental reports~\cite{wachs2020routine}.

\ctriangle{cyan} \textbf{Interview Data}~\cite{Anggraeny2023, Dorasamy_etal_2021, Fernandes_et_al_2023, kloess2019case, Chiu2022onlinegrooming, christiansen2024hidden, khotimah2024child, reneses2024he}: Interviews are widely used to collect structured demographic information and open-ended responses on cybergrooming experiences, offering rich qualitative insights. For example, \citet{Dorasamy_etal_2021} conducted interviews on social media use, sex education, and grooming risks, while \citet{Fernandes_et_al_2023} interviewed law enforcement officers to explore investigative challenges. To enhance credibility, some studies apply \textit{investigator triangulation}, where multiple researchers analyze transcripts independently to reduce subjective bias \cite{Fernandes_et_al_2023}. Additionally, member checking—where participants verify their responses—helps validate qualitative findings.

\begin{table}[t]
\small
\centering
\caption{Datasets Used in the Reviewed Social Science Cybergrooming Research}
\label{tab:datasets-social}
\vspace{-3mm}
\begin{tabular}{p{3cm} p{5cm} p{5cm} P{1cm}}
\toprule
\textbf{Source} & \multicolumn{1}{c}{\textbf{Dataset Content}} & \multicolumn{1}{c}{\textbf{Dataset Description}} & \textbf{Year}\\ 
\hline

\textbf{Childhood Online Sexual Abuse in the US} & 
Self-reported experiences of online sexual abuse from individuals aged 18–28, covering 11 abuse types. Data collected from 2,639 participants. & 
Survey-based dataset assessing the prevalence and dynamics of online sexual abuse, including grooming and image-based abuse, to inform prevention strategies. & 
2021 \textcolor{red}{$\blacktriangle$} \\
\multicolumn{4}{l}{\textbf{Link:} \url{https://www.icpsr.umich.edu/web/pages/index.html}}  \\
\midrule
\textbf{Online Child Sexual Abuse in Bangladesh} & 
Binary and continuous variables on internet use, online activity, and exposure to risks. & 
Mixed-method study with survey and interview data from grade 9–10 students in urban and rural schools. & 
2023 \textcolor{red}{$\blacktriangle$} \\
\multicolumn{4}{l}{\textbf{Link:} \url{https://data.mendeley.com/datasets/t8tm3ygfff/1}} \\
\bottomrule
\end{tabular}
\vspace{-5mm}
\end{table}

\ctriangle{lime} \textbf{Perverted Justice (PJ) Dataset}~\cite{black2015linguistic, Lorenzo-Dus_Izura_2017, van2016behavioural, lorenzo2016understanding, thomas2023offenders, lorenzo2020modus}: The PJ dataset~\cite{PJ-dataset}, consisting of chat logs between predators and decoys, is a key resource for analyzing grooming tactics and has been used to examine manipulation strategies, such as the strategic use of compliments to build trust~\cite{Lorenzo-Dus_Izura_2017}. However, reliance on decoy interactions may not fully capture authentic victim–predator dynamics. Some studies address this limitation by integrating PJ logs with victim narratives from legal case reports~\cite{lorenzo2016understanding}, thereby strengthening external validity.

\ctriangle{pink} \textbf{Document Analysis}~\cite{Anggraeny2023, Fernandes_et_al_2023, craven_current_2007, shannon2008sweden, seymour2021discursive}: Qualitative studies often use document analysis of books, academic literature, theses, and legal documents to complement primary data sources. For instance, \citet{Fernandes_et_al_2023} combined legal document analysis with interview data to construct a comprehensive framework for cybergrooming investigations. \citet{Anggraeny2023} analyzed child protection laws alongside offender interviews to assess how legal frameworks shape offender behaviors. While document analysis helps contextualize findings, it often depends on secondary sources, which may be outdated or lack direct relevance to current grooming behaviors.

Although these datasets provide valuable insights into cybergrooming, each has inherent limitations. Questionnaire and interview data rely on self-reports, which are subject to recall bias, social desirability, and underreporting. To improve reliability, some studies use \textit{triangulation}, combining interviews with multiple data sources~\cite{khotimah2024child}. Longitudinal datasets enable tracking behavioral change but often face participant attrition. The PJ dataset, while widely used, may not reflect authentic predator–victim dynamics due to its decoy structure. Document analysis informs legal and policy contexts but typically lacks direct empirical validation.

Table~\ref{tab:datasets-social} presents datasets commonly used in social science cybergrooming research, primarily relying on self-reported surveys and mixed-method studies to analyze victim experiences, risk factors, and online grooming prevalence.

These limitations highlight the need for more diverse, longitudinal, and context-rich datasets integrating multiple evidence sources. Future research should prioritize mixed-method approaches combining large-scale surveys, qualitative insights, case studies, and technological monitoring tools to enhance cybergrooming research depth and accuracy.

\subsection{Methods to Mitigate the Risk of Bias in Social Science Cybergrooming Research} \label{subsec:ss-methods-mitigate-risk-bias}

Biases in social science research on cybergrooming can affect study validity and reliability. To improve methodological rigor, researchers employ various mitigation strategies, each with strengths and limitations. Below, we discuss key biases and best practices to minimize their impact.

First, \textbf{\em experimental design biases}~\cite{Calvete_Orue_GamezGuadi_2022, Dorasamy_etal_2021, Wachs2012} arise from flaws in study design that influence outcomes. Best practices include {\em double-blind, randomized trials}~\cite{Calvete_Orue_GamezGuadi_2022}, where neither experimenters nor participants know the treatment assignment, reducing researcher influence. \citet{Calvete_Orue_GamezGuadi_2022} used this approach to assess an educational intervention on cybergrooming awareness. Another key strategy is {\em pre-tests}~\cite{Wachs2012} to identify design flaws before full data collection. \citet{Wachs2012} refined their questionnaire through pilot testing for clarity and reliability. Ensuring \textit{transparent participant selection}~\cite{Dorasamy_etal_2021} also prevents selection bias. \citet{Dorasamy_etal_2021} adopted a first-come, first-served approach in interviews to avoid preferential sampling. However, achieving fully neutral studies remains challenging, as transparency protocols must be rigorously maintained to prevent unintended biases.

Second, \textbf{\em missing data biases}~\cite{Finkelhor22-online-sexual-offense, HERNANDEZ2021106569} occur when participants leave survey items incomplete, leading to gaps in datasets. Common best practices include {\em Full Information Maximum Likelihood (FIML)}~\cite{HERNANDEZ2021106569, schoeps2020risk} and {\em data weighting}~\cite{Finkelhor22-online-sexual-offense} to estimate missing responses and adjust survey results. \citet{HERNANDEZ2021106569} applied FIML in their structural models to address missing responses, while \citet{Finkelhor22-online-sexual-offense} used weighted adjustments to improve result reliability. Another strategy involves {\em multiple imputation}, which replaces missing values with statistically plausible estimates, reducing bias from systematic data loss. Despite these methods, missing data remain a challenge, particularly when non-random dropout patterns introduce systematic errors. To further improve data completeness, researchers should adopt proactive approaches, such as conducting follow-up surveys and incentivizing full participation.

Third, \textbf{\em measurement biases}~\cite{Lorenzo-Dus_Izura_2017, wachs2020routine, black2015linguistic} arise from inaccuracies in data collection tools or procedures, affecting study reliability. Best practices include {\em inter-coder reliability checks}~\cite{Lorenzo-Dus_Izura_2017, black2015linguistic}, where multiple coders independently assess qualitative data for consistency. In multilingual research, a {\em standardized translation and back-translation process}~\cite{wachs2020routine} ensures conceptual consistency across languages. \citet{wachs2020routine} implemented a rigorous approach involving translation, independent back-translation, and version comparison to enhance accuracy. However, biases persist due to cultural nuances, varying survey interpretations, and self-report inaccuracies. To mitigate this, researchers should apply \textit{triangulation}, integrating multiple data sources (e.g., surveys, behavioral logs, and interviews) to cross-validate findings.

Finally, \textbf{\em sampling biases}~\cite{gamez2023stability, HERNANDEZ2021106569, Lorenzo-Dus_Izura_2017, thomas2023offenders} occur when certain groups are over- or underrepresented, limiting generalizability. Best practices include {\em random sampling}~\cite{gamez2023stability, HERNANDEZ2021106569, Lorenzo-Dus_Izura_2017}, giving all individuals in the target population an equal chance of selection. \citet{gamez2023stability} used a randomized approach to select 37 schools in Spain, ensuring regional diversity. \textit{stratified random sampling} improves representativeness by dividing participants into subgroups (e.g., age, gender, socioeconomic status) before selection. \citet{Lorenzo-Dus_Izura_2017} combined stratified and random sampling for a balanced PJ dataset. However, demographic imbalances and participant accessibility can still introduce bias. To address this, researchers should consider \textit{quota sampling} to maintain proportional demographic representation.

In summary, while these mitigation strategies significantly enhance research validity, adopting \textbf{\em best practices such as triangulation, stratified sampling, pre-testing, and standardized translation} can further reduce biases. Future studies should prioritize culturally adaptable research designs, integrate diverse data sources, and employ advanced statistical techniques to ensure the robustness and generalizability of cybergrooming research findings.

\subsection{Summary of Key Findings in Social Science Cybergrooming Research} \label{subsec:ss-summary-key-findings}

Our comprehensive review of cybergrooming studies in social sciences revealed several critical findings, highlighting actionable insights for policymakers, educators, and practitioners.

First, \textbf{\em education programs are essential} for effective cybergrooming prevention, as they increase awareness of risks and preventive strategies, significantly reducing incidents~\cite{Calvete_Orue_GamezGuadi_2022, whittle2014their, lundrigan2025relationship}. These programs equip children with the skills to recognize and avoid risky online behaviors, emphasizing the need for widespread implementation. Policymakers should prioritize mandatory digital safety education in school curricula, ensuring children are taught how to identify and respond to grooming attempts from an early age.

Second, \textbf{\em family protection plays a key role} in preventing cybergrooming, with open communication between parents, caregivers, and children crucial~\cite{whittle2014their}. Active parental involvement in internet use and early safety education fosters a safer environment, encouraging children to share online experiences. Research shows that perceived parental supervision discourages grooming attempts and strengthens defense against predators~\cite{kamar2022parental}. Practitioners should promote parental training programs that teach effective internet monitoring and child-focused communication strategies.

Third, \textbf{\em training for adults working with children} is crucial, as many professionals, including teachers and child protection workers, are motivated to prevent cybergrooming but lack specific training on this type of abuse~\cite{carmo2023knowledge}. Targeted training on grooming signs and tactics would enable these adults to act as early responders, detecting warning signs more accurately and intervening effectively before grooming escalates. Governments should invest in professional development programs that integrate cybergrooming awareness into teacher and childcare training modules.

Fourth, \textbf{\em gender differences shape} cybergrooming dynamics, highlighting the need for targeted prevention and intervention strategies~\cite{Finkelhor22-online-sexual-offense, HERNANDEZ2021106569, Wachs2012, Weingraber2020, gamez2021unraveling}. Grooming tactics and victim vulnerability vary by gender, with certain strategies more prevalent for specific groups. Prevention efforts should address gender-specific risk factors and tailor awareness campaigns to distinct grooming patterns across populations.

Fifth, \textbf{\em identifying risk factors for potential victims} is vital for designing targeted prevention measures~\cite{Finkelhor22-online-sexual-offense, gamez2023stability, HERNANDEZ2021106569, wachs2020routine, Wachs2012, Weingraber2020}. Awareness of \textbf{\em common grooming strategies}, such as praise~\cite{Anggraeny2023, Lorenzo-Dus_Izura_2017, gamez2018persuasion, black2015linguistic}, deception, gift-giving, and sexualization, is essential for developing countermeasures~\cite{gamez2021unraveling}. Online platforms should integrate AI-driven content moderation tools that flag grooming tactics in real time, helping to protect minors from predatory behaviors. \new{Moreover, recent evidence suggests that protective factors reduce grooming risk primarily by shaping adolescents’ online behavior patterns rather than exerting direct protective effects~\cite{aktu2024self}.}

Sixth, \textbf{\em online grooming strategies differ from offline methods}, with offenders often pretending to be younger to reduce perceived threat~\cite{gamez2018persuasion, oconnell2003typology}. Offenders assess risks early and exploit vulnerabilities from the start of interactions~\cite{black2015linguistic, shannon2008sweden}, presenting unique challenges for online protection. Law enforcement agencies should develop specialized units trained to detect and respond to digital grooming tactics, using proactive monitoring and AI-driven forensic analysis.

Seventh, \textbf{\em grooming tactics vary by gender}, with offenders engaging in lengthy emotional conversations with girls to build trust, while boys often face more direct, explicit chats involving age deception~\cite{van2016behavioural, seymour2021discursive}. These differing approaches suggest tailored interventions are necessary, with particular attention to how groomers manipulate trust in one group and exploit vulnerability more directly in the other. Digital platforms should develop algorithmic detection models that recognize gender-specific grooming tactics, enabling more precise intervention strategies.

Finally, \new{Peer/sibling victimization and prior sexual victimization consistently emerge as strong risk predictors, underscoring the importance of early identification and targeted prevention. As online engagement intensifies, policy responses must prioritize digital safety measures that mitigate minors’ exposure to exploitation.}

These findings underscore the urgent need for a multi-stakeholder approach involving policymakers, educators, parents, technology companies, and law enforcement agencies. Future research should focus on culturally adaptive prevention strategies, real-time detection systems, and cross-border collaboration to combat cybergrooming effectively.

\subsection{Limitations \& Lessons Learned in Social Science Cybergrooming Research} \label{subsec:ss-discussions}

Our comprehensive review of cybergrooming studies in social sciences revealed several limitations and key lessons learned. One major limitation is the \textbf{\em over-reliance on self-reported data}, which can introduce recall bias, social desirability bias, or underreporting, particularly in studies on sensitive topics like cybergrooming~\cite{almeida2024online, black2015linguistic, Calvete_Orue_GamezGuadi_2022, Finkelhor22-online-sexual-offense, gamez2023stability, HERNANDEZ2021106569, Weingraber2020, wachs2016cross, quayle2014rapid, gamez2018persuasion, christiansen2024hidden, almeida2025prevalence}. \textbf{\em Privacy concerns} further complicate data collection, particularly when involving minors, as ethical restrictions limit researchers’ access to critical but sensitive information~\cite{Dorasamy_etal_2021, Fernandes_et_al_2023, Lorenzo-Dus_Izura_2017, Wachs2012, van2016behavioural, evans2025corpus}. To address these challenges, future research should incorporate \textit{behavioral logging methods} and \textit{secure, anonymized data collection platforms} to improve data accuracy while ensuring ethical compliance.

Another key limitation is the \textbf{\em limited generalizability} of findings due to small sample sizes and reliance on potentially outdated data, restricting applicability to broader populations~\cite{Anggraeny2023, Dorasamy_etal_2021, Fernandes_et_al_2023, Finkelhor22-online-sexual-offense, gamez2023stability, HERNANDEZ2021106569, wachs2020routine, Weingraber2020, whittle2014their, wachs2016cross, van2016behavioural, gamez2018persuasion, kloess2019case, carmo2023knowledge, kamar2024relevance, Chiu2022onlinegrooming, christiansen2024hidden}. Future research should prioritize \textit{multi-country, large-scale studies} that include diverse age groups and socioeconomic backgrounds to enhance external validity. Additionally, longitudinal studies can track behavioral changes over time, providing deeper insights into grooming dynamics.

\textbf{\em Cultural and demographic gaps} remain a significant issue, as much of the existing research is conducted in Western contexts, limiting the understanding of cybergrooming risks in other cultural settings~\cite{wachs2016cross, villacampa2017online, melo2025grooming}. Additionally, the \textbf{\em underrepresentation of sexual minorities} highlights a critical gap, as these groups may experience distinct grooming patterns that current research fails to capture~\cite{Calvete_Orue_GamezGuadi_2022, Wachs2012, gamez2023stability}. Future studies should employ \textit{intersectional approaches} that examine how factors such as gender identity, ethnicity, and cultural norms influence grooming susceptibility. Expanding research to underrepresented regions, including low- and middle-income countries, would improve global understanding and inform more inclusive prevention strategies.

The findings highlight key priorities for strengthening prevention and intervention strategies. There is broad consensus on the need for \textbf{\em educational programs} to raise awareness of cybergrooming risks, as structured digital literacy initiatives have been shown to reduce victimization rates~\cite{Dorasamy_etal_2021, wachs2020routine, wachs2016cross, villacampa2017online}.  \new{However, exposure to campaign materials does not consistently increase disclosure of grooming attempts, which appears more strongly associated with relational factors such as family ties and parental openness~\cite{lundrigan2025relationship, aktu2024self}.} \new{Moreover, some awareness programs rely on stereotypical assumptions and lack empirical grounding, suggesting that prevention efforts should address the relational and communicative complexity of grooming interactions rather than simply encouraging refusal~\cite{reneses2024he, lorenzo5280035not}.} \textbf{\em Early education and parental involvement} are particularly crucial, as open discussions between parents and adolescents can foster safer online behaviors and increase reporting of suspicious interactions~\cite{Dorasamy_etal_2021, wachs2016cross}. 

Providing targeted training for professionals who work with children, such as teachers and social workers, could enable them to act as \textbf{\em first responders} in cybergrooming cases~\cite{carmo2023knowledge}. Future policies should encourage mandatory cybergrooming training for educators and youth service providers, equipping them with the tools to detect and intervene in early-stage grooming interactions. Additionally, tailored programs that consider \textbf{\em cultural differences} and \textbf{\em gender dynamics} are essential, as these factors influence the effectiveness of prevention efforts~\cite{wachs2016cross, villacampa2017online}. Governments and international organizations should support research collaborations to develop culturally adaptive intervention models.

Another lesson learned is the sophistication of \textbf{\em grooming tactics}, including deception, emotional manipulation, and trust-building strategies. Many grooming interactions now occur across multiple digital platforms, making detection more challenging~\cite{gamez2021unraveling}. Researchers and policymakers should advocate for \textit{enhanced AI-driven content monitoring tools} that can detect early grooming patterns and alert moderators in real time. Collaborations between academic institutions and tech companies could facilitate the development of automated tools that detect cross-platform grooming behaviors.

These limitations and lessons highlight the urgent need for \textbf{\em culturally inclusive, longitudinal, and multi-disciplinary approaches} in cybergrooming research. Future studies should prioritize diverse population samples, leverage emerging technologies for data collection and analysis, and strengthen collaborations between researchers, educators, and policymakers to enhance prevention, detection, and response strategies.

\section{Cybergrooming Research in Computational Sciences} \label{sec:cybergrooming-CS}

Over the past decade, cybergrooming has attracted increasing attention in computational sciences, leading to the development of technological solutions aimed at identifying and mitigating online grooming threats. Unlike social science research, which focuses on behavioral risk factors, victim experiences, and intervention strategies, computational studies primarily address detection through machine learning (ML), natural language processing (NLP), and data mining techniques. The emphasis on detection stems from the urgent need to automate identification on digital platforms, where vast volumes of online interactions make manual monitoring impractical. This section reviews computational studies, highlighting key research objectives, methods, datasets, and challenges.

\subsection{Research Objectives in Computational Science Cybergrooming Research} \label{subsec:cs-research-objectives}

Computational approaches to cybergrooming research center on four primary objectives:

\cbox{green!60!black} \textbf{Detecting Cybergrooming}~\cite{kim2020analysis, Ashcroft_Kaati_Meyer_2015, bours2019detection, sulaiman2019classification, pranoto2015logistic, bogdanova2014exploring, cano2014detecting, isaza2022classifying, amer2021detection, Borj_Bours_2019, Eilifsen_Shrestha_Bours_2023, fauzi2020ensemble, Munoz_Isaza_Castillo_2021, cook2023protecting, rezaee2023detecting, Gunawan16, Lykousas_Patsakis_2022, gupta2012characterizing, anderson2019intelligent, elzinga2012analyzing, ringenberg2024assessing, ALKHATEEB201614, zuo2018grooming, michalopoulos2010towards, milon2022take, waezi2024osprey, amuchi2012identifying, fauzi2023identifying, Michalopoulos14-gars, prosser2024helpful, mujtaba2025edgeaiguard, daish2024towards}: Most computational research focuses on detecting grooming behaviors, using ML and NLP to analyze chat logs for patterns of predatory behavior. Sections~\ref{subsec:cs-features} and~\ref{subsec:cs-eval} discuss detection techniques in greater detail.

\cbox{blue!60} \textbf{Educating on Cybergrooming}~\cite{Wang21-eancs, guo23-is, kim2023ai, rita2021chatbot, kupcova2024innovative, roumelioti2025leveraging, mateo2025groombuster}: Some studies aim to develop educational tools that raise awareness and promote prevention. \citet{Wang21-eancs} introduced SERI, a chatbot that simulates grooming conversations to educate users, while \citet{rita2021chatbot} created an interactive website providing resources and reporting mechanisms. \citet{kim2023ai} designed an AI-based simulation to enhance children's cybersecurity awareness. \new{In addition, gamified and curriculum-based platforms have been proposed, including a web application~\cite{kupcova2024innovative} and Cesagram for children~\cite{roumelioti2025leveraging}.}

\cbox{red!70} \textbf{Identifying Grooming Stages}~\cite{cano2014detecting, oconnell2003typology, elzinga2012analyzing, farag2023enhanced}: Understanding how grooming unfolds over time is another research goal. \citet{oconnell2003typology} proposed a six-stage grooming model, which has been leveraged in ML-based analyses~\cite{cano2014detecting}, applying machine learning to detect these stages in chat logs. \citet{farag2023enhanced} used the Luring Communication Theory Model (LCTM)~\cite{olson2007entrapping} to classify grooming patterns.

\cbox{yellow!90!} \textbf{Evaluating Victim Traits}~\cite{guo2023text}: A smaller subset of research examines victim behavior in online interactions. \citet{guo2023text} utilized text-mining techniques such as Linguistic Inquiry and Word Count (LIWC) to analyze chat data and identify social-psychological traits of potential victims.

Table~\ref{tab:research-objects-cs} summarizes the distribution of research objectives. The dominance of detection-based studies \new{(72.73\%)} highlights the field’s focus on automating cybergrooming identification, likely due to the scalability of ML models in processing vast amounts of chat data. \new{In contrast, educational interventions constitute only \new{15.91\%} of studies, and stage-based analysis accounts for just \new{9.09\%}, while research on victim traits remains notably scarce \new{(2.27\%)}.}

This emphasis on detection reflects an urgent need to equip platforms with real-time monitoring tools to prevent grooming before escalation. However, the limited focus on education, behavioral analysis, and victim profiling suggests critical gaps in proactive prevention strategies. Unlike social science research, which examines root causes and social interventions, computational approaches remain largely reactive, prioritizing post-interaction detection rather than preemptive intervention. Future research should explore integrating \textbf{\em detection with prevention}, incorporating ML-driven risk assessments into educational platforms, and leveraging AI-generated conversational simulations to prepare children and caregivers for identifying early grooming tactics. Expanding research on victim vulnerability factors can also improve predictive modeling, allowing for more targeted intervention strategies.

To create a more comprehensive approach to cybergrooming prevention and mitigation, future computational research should balance detection efforts with investment in \textbf{\em prevention, intervention, and victim behavior analysis}. Collaborations between computational scientists and social scientists can help bridge this gap, ensuring that technical solutions align with behavioral insights to maximize impact.

\begin{table}[htbp]
\small 
    \caption{Research Objectives in Computational Sciences for Cybergrooming} 
    \label{tab:research-objects-cs}
    \vspace{-3mm}
    \begin{tabular}{p{5cm}P{2.5cm}P{2.5cm}}
    \hline
         \multicolumn{1}{c}{\textbf{Research Objectives}} & \textbf{Percentage} & \textbf{Number of Articles} \\
    \hline
        \raisebox{-0.05cm}{\cbox{green!60!black}} Detecting Cybergrooming & \new{72.73\%} & \new{32} \\
    \hline
        \raisebox{-0.05cm}{\cbox{blue!60}}  Educating on Cybergrooming & \new{15.91\%} & \new{7} \\
    \hline
        \raisebox{-0.05cm}{\cbox{red!70}} Identifying Grooming Stages & \new{9.09\%} & \new{4} \\
    \hline
        \raisebox{-0.05cm}{\cbox{yellow!90}} Evaluating Victim Traits & \new{2.27\%} & \new{1} \\
    \hline
    \end{tabular}
    \vspace{1mm}
    
    \small{(Note: An article may appear multiple times if it addresses multiple objectives, each counted individually.)}
    \vspace{-5mm}
\end{table}

\subsection{Key Features of Cybergrooming Used in Computational Science Cybergrooming Research} \label{subsec:cs-features}

Developing detection algorithms to identify grooming activities is central to cybergrooming research. Key features used in grooming detection include:

\begin{itemize}
    \item \textbf{Content Features}: Analyzing shared text material, including {\em content patterns}~\cite{cano2014detecting, farag2023enhanced} for text complexity, readability, and sentence length; {\em discourse patterns}~\cite{cano2014detecting} to assess idea flow and semantic profiling of predatory behaviors; specific {\em grooming characteristics}~\cite{Ashcroft_Kaati_Meyer_2015, Gunawan16, cook2023protecting} to identify grooming behaviors; and tracking {\em frequency of images and URLs}~\cite{Ashcroft_Kaati_Meyer_2015} for external content sharing. These features are widely applied in industry settings, such as automated moderation tools in social media platforms and chat-based detection systems used by law enforcement agencies to flag potential grooming interactions in real time.

    \item \textbf{Emotional Features}: Capturing tone and emotions to reveal hidden intentions, including {\em emoticons}~\cite{Ashcroft_Kaati_Meyer_2015, Borj_Bours_2019} used to express emotions in chat interactions and {\em sentiment analysis}~\cite{bogdanova2014exploring, cano2014detecting} to detect sentiments like joy, sadness, or anger. These techniques have been integrated into AI-driven monitoring systems, such as child protection software that alerts parents or moderators when concerning sentiment patterns emerge in conversations.

    \item \textbf{Psychological Features}: Reflecting personality traits that signal grooming tendencies, including {\em psycho-linguistic patterns}~\cite{Borj_Bours_2019, cano2014detecting, gupta2012characterizing, kim2023ai} (often using LIWC for emotional and cognitive cues) and {\em psychological traits}~\cite{bogdanova2014exploring}, such as neuroticism, which may indicate predatory behavior. Law enforcement agencies utilize these analyses in forensic cybercrime investigations, where suspect chat histories are examined for manipulation tactics and personality-driven behavioral patterns.

    \item \textbf{Textual Features}: Derived from word frequency, phrases, structure, and style, textual techniques include Bag of Words for analyzing word and character patterns~\cite{bogdanova2014exploring, Borj_Bours_2019, bours2019detection, cano2014detecting, fauzi2020ensemble, Munoz_Isaza_Castillo_2021}; TF-IDF (Term Frequency-Inverse Document Frequency) with ML classifiers for identifying predators in chats~\cite{Borj_Bours_2019, Eilifsen_Shrestha_Bours_2023, fauzi2020ensemble, Gunawan16, pranoto2015logistic, anderson2019intelligent, farag2023enhanced}; Part of Speech analysis to detect predators’ lexical behaviors~\cite{Ashcroft_Kaati_Meyer_2015, cano2014detecting, fauzi2023identifying}; and TF with ML algorithms to measure suspect word frequency~\cite{Eilifsen_Shrestha_Bours_2023, fauzi2020ensemble}. These features are extensively used in industry-grade cyber defense tools, such as AI-powered content moderation systems deployed by social media companies to automatically filter predatory content in messaging platforms.

\end{itemize}

\textbf{\em Real-World Applications}: These features have enabled practical cybergrooming detection tools. Law enforcement agencies like the FBI and INTERPOL use ML-based systems to analyze chat data and identify grooming. Social media platforms such as Facebook, Instagram, and Discord deploy NLP-based moderation to detect and prevent grooming conversations, enabling faster intervention. Additionally, nonprofit organizations and educational platforms use AI-driven chatbots to educate children about grooming risks through interactive simulations with real-time detection.

\textbf{\em Limitations and Future Directions}: While cybergrooming detection research has made significant strides, notable challenges remain. {\em Generalizability across languages and cultures} is a key limitation, as most NLP models are trained on English-language datasets and may not accurately detect grooming behaviors in other linguistic or cultural contexts. Grooming tactics vary across regions, and certain conversational nuances may not be effectively captured by current models, leading to false positives or missed detections. Future research should focus on developing {\em multilingual and culturally adaptive models}, ensuring that detection techniques are robust across diverse populations.

Another limitation is the reliance on {\em text-based analysis}, which may overlook crucial non-verbal cues such as voice tone, video behaviors, or even behavioral metadata (e.g., frequency and time of messages). Addressing this gap requires integrating {\em multimodal detection approaches} that incorporate video, audio, and behavioral analytics into cybergrooming detection frameworks. Collaboration between researchers, industry professionals, and policymakers is essential to refine detection models and enhance the effectiveness of digital safety measures.

These insights emphasize the importance of balancing technological detection with ethical and practical considerations. Future efforts should explore {\em privacy-preserving AI models} that respect user data rights while improving the accuracy and effectiveness of cybergrooming detection across global digital platforms.

\subsection{Evaluation Methods and Metrics in Computational Science Cybergrooming Research}\label{subsec:cs-eval}

\new{Across the reviewed computational literature, evaluation practices follow recurring patterns. Most studies emphasize classification performance on annotated datasets, while fewer address longitudinal modeling, victim-centered outcomes, or real-time intervention. Common limitations include dataset bias, language and cultural constraints, and trade-offs between model complexity and interpretability.}

Researchers in computational sciences studying cybergrooming use various machine learning (ML), text-mining, and natural language processing (NLP) techniques to develop detection tools. ML models, particularly those using NLP, dominate due to the \textbf{\em text-heavy nature} of grooming interactions.  Techniques like {\em tokenization} and {\em sentiment analysis} aid detection by capturing nuanced textual details but require {\em high contextual accuracy} and {\em quality data}, posing challenges in handling {\em cultural nuances} and {\em implicit meanings} that demand careful model tuning. \citet{amer2021detection} employed word embeddings and pre-trained models like GloVe for effective text feature extraction, enhancing detection accuracy. \citet{anderson2019intelligent} used the {\em bag of words (BoW)} method to identify key lexical features in online child grooming texts. \citet{bogdanova2014exploring} examined high-level feature extraction techniques, such as {\em sentiment analysis} and {\em parsing}, to analyze textual features reflecting emotional and psychological states, identifying predatory behavior.

Beyond traditional NLP techniques, advanced models like \textbf{\em Support Vector Machines (SVMs)} and \textbf{\em Convolutional Neural Networks (CNNs)} have been applied. \citet{Gunawan16} deployed SVM alongside {\em k-nearest neighbors (KNN)} to detect online child grooming conversations, achieving high accuracy in classifying conversations based on predefined grooming characteristics. \citet{rezaee2023detecting} explored {\em contrastive chat embeddings} to enhance the effectiveness of SVMs by distinguishing typical from atypical conversation patterns. Deep learning models such as CNNs have been leveraged for their ability to detect complex patterns in unstructured text data. \citet{Munoz_Isaza_Castillo_2021} used CNNs to extract invariant features across chat logs, highlighting their utility in monitoring online interactions. However, CNNs require {\em large, annotated datasets}, which are often costly and scarce. Simpler models, such as {\em Naïve Bayes}, remain relevant for their {\em ease of interpretation}, although they may oversimplify linguistic dependencies. 

\new{A central limitation across computational studies is \textbf{\em dataset composition}, as many benchmark corpora remain predator-centric. This bias leads models to excel at identifying offenders while struggling to detect vulnerable victims.} \citet{guo2023text} highlighted this limitation, showing that current models classify groomer intent more accurately than victim susceptibility. Additionally, {\em linguistic and cultural biases} in predominantly English-language datasets reduce accuracy across different languages and contexts. To address these issues, researchers should adopt {\em data augmentation techniques} and develop {\em multilingual training corpora} to improve model generalizability\new{; a structured overview of these evaluation methods is provided in Table~\ref{tab:evaluation-methods-computational} in Appendices.}

To evaluate these tools, studies employ various metrics, summarized in \new{Table~\ref{tab:evaluation-metrics-computational} in Appendices.} Key metrics include {\em precision and recall}, which are vital for detection accuracy. Precision minimizes false positives, essential in social and legal contexts, while recall ensures comprehensive case coverage. \new{The {\em F-Score} balances both precision and recall for evaluating a single model’s performance, but can be affected by data imbalance. 
To complement this, studies also report {\em ROC curves and AUROC scores}, which assess a model’s discrimination ability across thresholds and are particularly useful for comparing models or handling imbalanced datasets.} Additionally, \citet{Eilifsen_Shrestha_Bours_2023} utilized {\em Youden’s Index} to optimize sensitivity and specificity, ensuring effective differentiation between grooming and non-grooming conversations. 

\textbf{\em Human evaluation} is also employed in computational cybergrooming research. \citet{Wang21-eancs} randomly selected 200 conversation samples to be evaluated by three human graders, who compared chatbot-generated responses to those from the PJ dataset, determining which response was more appropriate based on historical conversational context. This approach ensures that chatbot responses are both {\em realistic} and {\em appropriate for preventing cybergrooming}. 

\new{Beyond model-level assessment, human evaluation also informs prevention-oriented systems. \citet{kupcova2024innovative} evaluated a web application with elementary students, while \citet{roumelioti2025leveraging} employed multi-country workshops with pre- and post-assessments, and pilot user studies further demonstrate the educational effectiveness of narrative-based gameplay approaches~\cite{mateo2025groombuster}, collectively signaling a shift toward user-centered evaluation.}

Addressing these challenges requires a more \textbf{\em comprehensive, multidisciplinary approach}. Future research should focus on developing {\em multilingual NLP models} to improve detection across linguistic and cultural settings. Additionally, {\em integrating multimodal features}, such as behavioral metadata and voice sentiment analysis, could enhance model robustness beyond text-based analysis. Ethical considerations should also be prioritized, ensuring that AI-driven detection systems balance {\em privacy concerns} with effective intervention strategies. Collaboration between computational and social scientists is essential to refine detection frameworks and integrate them into real-world applications, bridging the gap between technological advancements and practical interventions.

\begin{table}[t]
\small 
\centering
\caption{Datasets Used in the Reviewed Computational Science Cybergrooming Research}
\label{tab:datasets-computational}
\vspace{-3mm}
\begin{tabular}{p{3cm} p{5cm} p{5cm} P{1cm}}
\toprule
\textbf{Source} & \multicolumn{1}{c}{\textbf{Dataset Content}} & \multicolumn{1}{c}{\textbf{Dataset Description}} & \textbf{Year}\\ 
\hline

\textbf{Literotica} & 
User-generated erotic literature and conversation scripts. & 
Online platform hosting 1.25 billion words of erotic content, attracting 2.6–3.1 million unique visitors annually. & 
1996 \textcolor{green}{$\blacktriangle$} \\
\multicolumn{4}{l}{\textbf{Link:} \url{https://literotica.com/}} \\
\hline

\textbf{NPS Chat} & 
Chat messages with metadata on dates, age groups, and part-of-speech/dialog-act annotations. & 
Chat-based dataset for studying language patterns and behaviors associated with grooming and online risks. & 
2006 \textcolor{orange}{$\blacktriangle$} \\
\multicolumn{4}{l}{\textbf{Link 1:} \url{https://www.kaggle.com/datasets/nltkdata/nps-chat}}\\
\multicolumn{4}{l}{\textbf{Link 2:} \url{https://catalog.ldc.upenn.edu/LDC2010T05}} \\
\hline

\textbf{Perverted Justice} & 
Chat logs in CSV format with metadata on conversation dates and user interactions. & 
Dataset containing decoy-predator chat logs used for exposing online predators. & 
2012 \textcolor{lime}{$\blacktriangle$} \\
\multicolumn{4}{l}{\textbf{Link 1:} \url{https://web.archive.org/web/20240127030926/}} \\ 
\multicolumn{4}{l}{\textbf{Link 2:} \url{http://www.perverted-justice.com/}}\\
\multicolumn{4}{l}{\textbf{Link 3:} \url{https://ieee-dataport.org/documents/curated-pj-dataset}} \\
\hline

\textbf{PAN 2012} & 
XML file with 60,000 training and 155,000 test chat logs, plus predator identification files. & 
Dataset from PAN 2012 competition containing predator and non-predator conversations from IRC, Omegle, and PJ datasets. & 
2012 \textcolor{brown}{$\blacktriangle$} \\
\multicolumn{4}{l}{\textbf{Link:} \url{https://zenodo.org/records/3713280}} \\
\hline

\textbf{PAN 2013} & 
Blog posts labeled with author demographics (age, gender) in English and Spanish. & 
Dataset from PAN 2013 competition for author profiling and linguistic analysis across demographic groups. & 
2013 \textcolor{black}{$\blacktriangle$} \\
\multicolumn{4}{l}{\textbf{Link:} \url{https://zenodo.org/records/3715864}} \\
\hline

\textbf{LiveMe SLSS} & 
39 million chat messages from 1.4 million users during 293,271 live broadcasts (2016–2018). & 
Dataset analyzing viewer-streamer interactions and potential grooming behaviors in live streams. & 
2018 \textcolor{violet}{$\blacktriangle$} \\
\multicolumn{4}{l}{\textbf{Link:} \url{https://zenodo.org/records/3560365}} \\
\hline

\textbf{ChatCoder} & 
Chat transcripts between convicted predators and adults posing as minors, with labeled metadata. & 
Annotated dataset for analyzing cyber-predatory behavior and cyberbullying patterns. & 
2020 \textcolor{yellow}{$\blacktriangle$} \\
\multicolumn{4}{l}{\textbf{Link:} \url{https://chatcoder.com/}} \\
\bottomrule
\end{tabular}
\vspace{-5mm}
\end{table}

\subsection{Evaluation Datasets in Computational Science Cybergrooming Research} \label{subsec:eval-datasets}
Computational science research utilizes several key datasets to detect cybergrooming in online communications. These datasets provide essential training and evaluation resources for machine learning models, enabling the identification of predatory behaviors in digital interactions.

\ctriangle{lime} \textbf{Perverted Justice (PJ)}~\cite{pranoto2015logistic, sulaiman2019classification, Ashcroft_Kaati_Meyer_2015, bogdanova2014exploring, bours2019detection, cano2014detecting, Gunawan16, guo2023text, gupta2012characterizing, Wang21-eancs, anderson2019intelligent, elzinga2012analyzing, cook2023protecting, rezaee2023detecting, milon2022take, prosser2024helpful}: Collected by the Perverted Justice organization, this dataset comprises chat logs where adult decoys interact with potential predators posing as minors. The dataset is stored in CSV format and contains rich metadata, facilitating the study of grooming behaviors and language patterns in online conversations~\cite{PJ-dataset}. Despite its extensive use, its reliance on decoy interactions may introduce biases not entirely reflective of real-world grooming tactics.

\ctriangle{brown} \textbf{PAN 2012}~\cite{amer2021detection, Ashcroft_Kaati_Meyer_2015, Borj_Bours_2019, bours2019detection, Eilifsen_Shrestha_Bours_2023, fauzi2020ensemble, isaza2022classifying, Munoz_Isaza_Castillo_2021, rezaee2023detecting, milon2022take}: The PAN 2012 dataset~\cite{inches2012pan12} integrates chat logs from PJ, IRC logs, and Omegle, offering a diverse set of conversations. It includes both predator and non-predator interactions, making it valuable for training machine learning models to distinguish grooming behaviors. However, its platform-specific content may limit generalizability across different online environments.

\ctriangle{black} \textbf{PAN 2013}~\cite{anderson2019intelligent}: Originally designed for age and gender profiling, the PAN 2013 dataset~\cite{PAN13} contains blog posts labeled with author demographics. While not explicitly intended for grooming detection, it is often combined with PJ data to enhance linguistic diversity and improve models’ ability to generalize across various forms of online communication.

\ctriangle{green} \textbf{Literotica}~\cite{pranoto2015logistic, sulaiman2019classification}: The Literotica dataset~\cite{literotica} comprises user-generated erotic stories, totaling over a billion words. It provides a resource for studying language patterns associated with grooming and predatory narratives. However, as it consists primarily of fictional content, its application to real-world grooming detection remains limited.

\ctriangle{yellow} \textbf{ChatCoder2}~\cite{kim2020analysis, waezi2024osprey, amuchi2012identifying, michalopoulos2010towards, milon2022take, Michalopoulos14-gars, prosser2024helpful}: Developed as part of the NSF-funded "Tracking Predators" project, ChatCoder2 includes annotated chat logs of convicted predators interacting with decoys~\cite{chatcoder2}. This dataset allows for in-depth behavioral analysis, providing structured metadata that facilitates the identification of grooming patterns. However, like PJ, its reliance on decoy interactions may not fully capture the complexity of real grooming behaviors.

\ctriangle{lightgray} \textbf{Experimental Results}~\cite{oconnell2003typology, raihana2024implementation, kupcova2024innovative, mateo2025groombuster}: These datasets stem from experimental setups where researchers simulate child-user interactions in chat environments to study grooming stages. While these controlled environments allow for targeted data collection, they may lack the spontaneity and variability of real-world grooming attempts.

\ctriangle{red} \textbf{Questionnaire Data}~\cite{rita2021chatbot, roumelioti2025leveraging, mateo2025groombuster}: Surveys and self-reported data provide insights into public awareness, chatbot effectiveness in cybergrooming detection, and user experiences with online safety~\cite{rita2021chatbot}. While valuable, self-reporting biases and limited sample sizes may affect the reliability and generalizability of findings.

\ctriangle{cyan} \new{\textbf{Interview Data}~\cite{daish2024towards}: Expert and end-user interviews reveal usability challenges in AI-assisted cybergrooming detection, particularly in interpreting classification scores and providing effective feedback for continuous learning~\cite{daish2024towards}. Such insights inform Human-in-the-Loop design refinement but remain limited by small samples and subjectivity.}

\ctriangle{violet} \textbf{LiveMe SLSS}~\cite{Lykousas_Patsakis_2022, fauzi2023identifying, michalopoulos2010towards, milon2022take, guo23-is, Michalopoulos14-gars}: The LiveMe SLSS dataset~\cite{lykousas2021large} contains over 39 million chat messages from LiveMe broadcasts, enabling analysis of real-time grooming behaviors and audience–streamer interactions. However, the live-stream data are often noisy and unstructured, complicating accurate annotation and analysis.

\ctriangle{orange} \textbf{NPS Chat}~\cite{bogdanova2014exploring, amuchi2012identifying}: The NPS Chat corpus~\cite{forsyth2010nps} contains chat room posts labeled with age information, part-of-speech tags, and dialogue acts. This dataset aids in understanding conversational patterns and linguistic cues in online interactions, but does not specifically focus on grooming detection.

Table~\ref{tab:datasets-computational} in Section~\ref{sec:datasets_twofields} summarizes these datasets. Computational cybergrooming research relies primarily on annotated chat logs to train and evaluate detection algorithms. Key datasets such as PJ, PAN~2012, and PAN~2013 provide realistic conversational data for developing and refining machine learning models, particularly in supervised settings where labeled interactions distinguish predatory from benign conversations. However, reliance on decoy-based interactions limits authenticity and may not fully capture the nuanced tactics of real predators.

Newer datasets, such as LiveMe SLSS and ChatCoder2, reflect a shift toward dynamic, real-world interactions. LiveMe SLSS focuses on live-streaming environments, introducing challenges related to real-time grooming detection and the evolving nature of grooming tactics. ChatCoder2, with its structured annotations, enhances behavioral analysis and facilitates model refinement. Despite their advantages, these datasets also present challenges, such as noise in live-streaming data and platform-specific biases that may limit generalizability.

In summary, computational cybergrooming datasets advance detection models but remain constrained by artificial setups, ethical concerns, and limited authenticity. Future research should prioritize more diverse, real-world data while maintaining strong ethical safeguards to better capture evolving grooming tactics across platforms.

\subsection{Methods to Mitigate the Risk of Bias in Computational Science Cybergrooming Research} \label{subsec:cs-methods-risk-bias}

Sampling bias and class imbalance bias are notable challenges in computational cybergrooming research. \textbf{\em Sampling bias} occurs when the sampling process disproportionately favors certain groups or outcomes, limiting generalizability~\cite{bogdanova2014exploring, Eilifsen_Shrestha_Bours_2023, gupta2012characterizing, Gunawan16, Munoz_Isaza_Castillo_2021, Wang21-eancs, amer2021detection}. To reduce this bias, researchers employ \textbf{\em random sampling} to select diverse subsets, improving representativeness. For instance, \citet{bogdanova2014exploring} randomly sampled 20 chat logs from the Cybersex and NPS chat corpora to build a testing dataset for detecting cyberpedophilia. \citet{Eilifsen_Shrestha_Bours_2023} implemented a \textbf{\em dynamic sampling method} that adjusts sliding window sizes based on conversation flow to capture a broader range of grooming behaviors. Similarly, \citet{Gunawan16} randomly selected grooming conversations from the PJ dataset and non-grooming conversations from Literotica to enhance model robustness. \citet{amer2021detection} applied \textbf{\em stratified sampling}, ensuring training data for artificial neural networks covered diverse sexual harassment scenarios and chat predator detection. These approaches minimize bias by creating datasets that better represent various online interactions.

\textbf{\em Class imbalance bias} arises when one class (e.g., predatory or non-predatory conversations) dominates, skewing model performance by favoring the majority class~\cite{Borj_Bours_2019, cano2014detecting, fauzi2020ensemble}. Studies address this issue using \textbf{\em fixed sampling}, \textbf{\em data augmentation}, and \textbf{\em synthetic data generation}. \citet{cano2014detecting} applied fixed sampling, adding a predefined number of samples, such as randomly selecting 3,304 sentences, to balance datasets. \new{\citet{mujtaba2025edgeaiguard} leveraged GPT-4 for data augmentation to enhance detection robustness under limited labeled data.} \citet{Borj_Bours_2019} leveraged the PAN12 dataset, which contains 64,911 regular and 2,016 predatory conversations, and attempted to reduce imbalance by selectively removing overrepresented conversations. Additionally, researchers integrate datasets such as ChatCoder2~\cite{kim2020analysis, milon2022take, Michalopoulos14-gars} and PAN2013~\cite{anderson2019intelligent} to enhance the representation of underrepresented classes. However, while these methods improve balance, they may oversimplify the complexity of real-world grooming interactions.

Models using \textbf{\em generative adversarial networks (GANs)} can generate synthetic grooming conversations that resemble real interactions, enriching training data and reducing overfitting to specific linguistic patterns. GAN-based data augmentation has improved classifier performance in domains such as fraud detection by addressing class imbalance~\cite{strelcenia2023survey}. Additionally, \textbf{\em reinforcement learning-based bias correction} can dynamically adjust decision boundaries, enhancing fairness and adaptability in classification systems~\cite{lewis2013reinforcement}.

This analysis indicates that random and fixed sampling improve data diversity and model generalization, though artificially balanced subsets may oversimplify real-world complexity. Future work should incorporate \textbf{\em adversarial training, GAN-based augmentation, and reinforcement learning strategies}~\cite{li2025novel} to refine class distributions while preserving conversational nuances. Integrating these methods with traditional sampling can enhance the fairness, accuracy, and adaptability of cybergrooming detection models.

\subsection{Summary of Key Findings in Computational Science Cybergrooming Research} \label{subsec:cs-summary-key-findings}

First, \textbf{\em high-level features} are crucial for detecting online grooming, including emotional markers, fixated discourse, and deceptive behavior, significantly improving accuracy in distinguishing grooming from other interactions~\cite{Ashcroft_Kaati_Meyer_2015, bogdanova2014exploring, kim2020analysis}. Computational research trends show a shift toward deep learning-based models, particularly recurrent neural networks (RNNs) and transformer architectures, which process sequential chat data with better contextual understanding. Advanced models, such as GRU and BiLSTM combinations, have achieved up to 99.12\% accuracy in harassment detection, underscoring their effectiveness in identifying harmful content~\cite{amer2021detection}. These findings highlight the increasing reliance on adaptive learning systems that evolve with grooming strategies.

Second, \textbf{\em context-aware detection systems} support early identification of predatory behaviors. These systems use real-time NLP and sequential pattern analysis to classify conversations as predatory or non-predatory~\cite{Ashcroft_Kaati_Meyer_2015, Borj_Bours_2019, Eilifsen_Shrestha_Bours_2023, bours2019detection, gupta2012characterizing, fauzi2020ensemble, Gunawan16, isaza2022classifying}. Visual tools, such as conversational heatmaps and time-series models, enable dynamic risk tracking and faster intervention~\cite{elzinga2012analyzing}. Integrating explainable AI (XAI) enhances transparency and trust.

Third, \textbf{\em human-machine collaboration} helps reduce bias and prevent over-prediction in ML-based detection of predatory communication. While ML approaches are efficient, they struggle to differentiate deceptive grooming tactics from benign interactions. Incorporating \textit{human-in-the-loop validation}, where annotators intervene in ambiguous cases, improves precision with minimal effort compared to full human annotation~\cite{cook2023protecting}. This approach ensures edge cases, such as context-dependent conversations, are accurately classified, reducing false positives and negatives. The shift toward \textit{hybrid AI-human moderation} aligns with industry trends, where social media monitoring and child safety applications integrate human oversight to refine detection models.

Fourth, \textbf{\em victim characteristics} significantly influence grooming vulnerability. Identity and personality traits affect susceptibility, while cognitive resilience and strong social support serve as protective factors~\cite{guo2023text}. Studies highlight predictive grooming behaviors, such as reframing conversations, soliciting explicit images, and escalating interactions to sexual content, aiding detection and forensic investigations~\cite{pranoto2015logistic}. Computational models reveal distinct grooming strategies when targeting real adolescents versus adult impersonators (e.g., undercover officers or chatbots). Differences in risk assessment, meeting arrangements, and conversational control suggest that groomers tailor tactics based on perceived victim profiles~\cite{ringenberg2024assessing}. This insight informs law enforcement, emphasizing the need for \textit{adaptive decoy models} that mimic adolescent behaviors to enhance grooming intervention success.

Fifth, \textbf{\em interactive platforms and chatbots} are effective in preventing grooming and facilitating abuse reporting, demonstrating their growing role in educational and real-time intervention systems~\cite{rita2021chatbot, Wang21-eancs}. AI-driven conversational agents are being developed to engage potential victims in simulated grooming interactions, helping them recognize risks before they escalate~\cite{raihana2024implementation}. \new{In addition to intervention, gamified educational platforms are emerging as a key trend in prevention efforts, enhancing children's awareness of grooming risks through interactive learning environments~\cite{roumelioti2025leveraging}.} \new{Detection interfaces are also evolving to support human-in-the-loop feedback, emphasizing simple confidence indicators and streamlined agree/disagree workflows~\cite{daish2024towards}.}

Finally, \textbf{\em relationship formation} is a key grooming phase, with groomers initially building trust before introducing sexual content. The latest computational research suggests that early-stage relational analysis should be prioritized in detection models, focusing on sentiment shifts, grooming-specific linguistic markers, and progression modeling to identify risk before escalation~\cite{cano2014detecting, gupta2012characterizing, Lykousas_Patsakis_2022}. Practical applications include preemptive alert systems for parents and educators, where AI models detect early grooming attempts and generate warnings before explicit content is introduced.

These findings reflect \textbf{\em a broader shift in computational cybergrooming research toward real-time prevention, hybrid human-AI collaboration, and explainable AI systems}. Future research should focus on enhancing cross-platform grooming detection, improving bias mitigation techniques, and integrating multilingual capabilities to ensure broader applicability across diverse digital environments.

\subsection{Limitations \& Lessons Learned in Computational Science Cybergrooming Research}

Our review of cybergrooming research highlights numerous limitations and critical lessons that underline the complexity of addressing grooming behavior in computational settings. Researchers emphasize the \textbf{\em importance of awareness}, advocating for educational initiatives in schools to help students and educators recognize cybergrooming risks and equip them with preventative strategies~\cite{oconnell2003typology, rita2021chatbot, Wang21-eancs}. However, a significant limitation lies in the \textbf{\em limited generalizability} of detection models, which are often trained on narrow, specific datasets, challenging their applicability in varied real-world scenarios~\cite{bours2019detection, Eilifsen_Shrestha_Bours_2023, fauzi2020ensemble, gupta2012characterizing, elzinga2012analyzing, waezi2024osprey, amuchi2012identifying, fauzi2023identifying, milon2022take, daish2024towards, mateo2025groombuster}. This limitation suggests a need for more \textbf{\em representative and diverse datasets} to capture the full scope of online interactions.

\textbf{\em Cross-cultural differences} further complicate detection efforts, as the interpretation and meaning of certain language patterns vary significantly across cultural contexts~\cite{rezaee2023detecting}. This calls for models adaptable to cultural nuances, which is essential in global contexts. Additionally, many computational approaches for cybergrooming detection demand \textbf{\em significant computing resources}, emphasizing the need for more resource-efficient algorithms, especially for use in low-resource environments where access to advanced hardware may be restricted~\cite{fauzi2020ensemble, Munoz_Isaza_Castillo_2021, mujtaba2025edgeaiguard}.

The \textbf{\em lack of high-quality datasets} remains a major barrier in this field, with data scarcity, privacy constraints, and poor readability affecting training quality~\cite{Ashcroft_Kaati_Meyer_2015, bogdanova2014exploring, Borj_Bours_2019, bours2019detection, Eilifsen_Shrestha_Bours_2023, guo2023text, gupta2012characterizing, Wang21-eancs, amer2021detection, isaza2022classifying, sulaiman2019classification, pranoto2015logistic, anderson2019intelligent}. Additionally, model adaptability across \textbf{\em languages and linguistic variations} remains a challenge, as models must handle slang, regional dialects, and common misspellings to maintain accuracy in diverse online communities~\cite{Ashcroft_Kaati_Meyer_2015, Munoz_Isaza_Castillo_2021, Wang21-eancs, pranoto2015logistic, guo23-is}.

A critical yet often overlooked challenge is the \textbf{\em ethical concerns in machine learning-based grooming detection}. False positives, where benign conversations are mistakenly flagged as grooming, pose serious risks, particularly in legal and social contexts. Wrongful accusations can lead to reputational damage and legal consequences. Conversely, false negatives, or missed detections, allow predatory behavior to persist, endangering victims. To address these risks, researchers should explore \textbf{\em explainable AI (XAI)} to enhance transparency in model decisions, improving human oversight. Incorporating \textbf{\em human-in-the-loop systems}, where moderators validate high-risk predictions, can mitigate false positives while maintaining efficiency.

To overcome these limitations, \textbf{\em interdisciplinary collaboration} is essential. Integrating insights from psychology, sociology, and linguistics can enhance model development and refine detection capabilities. Partnering with educators, law enforcement, and child protection agencies can foster a holistic approach, combining technology with social awareness to more effectively counteract cybergrooming~\cite{ALKHATEEB201614, Ashcroft_Kaati_Meyer_2015, guo2023text, gupta2012characterizing, Munoz_Isaza_Castillo_2021, rita2021chatbot}. Additionally, future research should focus on differential privacy techniques to ensure user data protection while improving detection reliability. This cross-disciplinary engagement can help address the unique complexities of grooming detection and support the development of robust, adaptable, and ethically responsible solutions. 

\section{Discussions}
\subsection{Multidisciplinary Approaches}

\new{Cybergrooming research has evolved along two largely parallel paths: social science emphasizes theory-driven, human-centered analyses of grooming processes and risk factors, while computational research focuses on scalable detection using machine learning and NLP. Interdisciplinary integration remains limited, though early convergence appears through shared datasets and complementary methods. Social science increasingly applies computational techniques to validate theoretical models~\cite{gamez2023stability, wachs2020routine}, while computational work draws on grooming theories~\cite{olson2007entrapping} to structure detection pipelines~\cite{farag2023enhanced}. More substantive integration is emerging in human-centered AI, where domain expertise is embedded into system design, such as iterative co-design frameworks that combine linguistic analysis, ML development, and law enforcement workflows through human-in-the-loop feedback~\cite{daish2024towards}.}

\new{Despite progress, integration remains partial and mostly unidirectional. Psychological and sociological constructs are often incorporated into computational systems in static ways, rarely shaping model architecture or adaptation, whereas computational methods in the social sciences are primarily used for analysis rather than for theory refinement. These gaps highlight the need for deeper, bidirectional alignment between theory and computational mechanisms. Emerging multidisciplinary efforts aim to move beyond parallel development by systematically integrating social science theories with detection, prevention, and policy systems, which can be grouped into three complementary directions.}

\new{\textbf{Stage-aware and vulnerability-informed modeling.}
Social science characterizations of grooming as a staged, dynamic process can inform stage-aware computational models that track conversational progression rather than isolated messages. Similarly, empirically validated vulnerability factors (e.g., emotional dependency, social isolation) can be operationalized as contextual signals to support adaptive risk assessment and targeted intervention.}

\new{\textbf{Human-in-the-loop and intervention-oriented systems.}
Multidisciplinary integration enables human-in-the-loop frameworks in which computational models perform large-scale screening while human experts provide contextual interpretation, ethical oversight, and feedback. This interaction supports behaviorally grounded detection and facilitates interventions such as personalized warnings, educational nudges, or escalation protocols.}

\new{\textbf{Dataset-level and cross-cultural integration.}
Multidisciplinary collaboration can mitigate Western-centric bias by systematically incorporating non-Western studies and leveraging theory-driven social science datasets from diverse cultural contexts for data augmentation or simulation in computational models. Integrating cross-cultural and multilingual perspectives helps capture variation in grooming strategies and communicative norms, supporting culturally sensitive prevention and globally applicable detection systems.}

\new{Together, these examples demonstrate how psychological theory can shape computational design and how computational scalability can extend socially grounded insights, enabling more adaptive, context-aware, and ethically informed cybergrooming detection and prevention systems.}

To encourage more interdisciplinary research, funding initiatives such as the National Science Foundation (NSF) Secure and Trustworthy Cyberspace (SaTC) program and the European Horizon Cybersecurity Research Initiative offer potential avenues for supporting joint projects. Establishing research centers focused on AI ethics in online safety or cross-disciplinary cybersecurity research hubs can further promote cooperation between social and computational scientists. By fostering collaboration through dedicated funding and institutional support, future cybergrooming detection strategies can become more adaptable, culturally informed, and ethically sound, strengthening global efforts in prevention, detection, and policy development.

\subsection{Answers to Research Questions}
\label{sec:discussions}

\textbf{RQ1.} \textbf{\em What are the key approaches used to study cybergrooming in social and computational sciences?}

Social science research employs \textbf{\em qualitative methods}, such as interviews and literature reviews, to investigate victim profiles, grooming tactics, and the psychological impacts of cybergrooming. \\ \vspace{-4mm}

These approaches provide deep contextual insights, uncovering perceptions and behaviors often missed by quantitative methods. \textbf{\em Quantitative methods}, including surveys and statistical analyses, enable large-scale data collection and pattern identification, clarifying risk factors and victim–offender dynamics across diverse populations.

\textbf{RQ2.} \textbf{\em What are the primary contributions of existing cybergrooming research in both fields?}

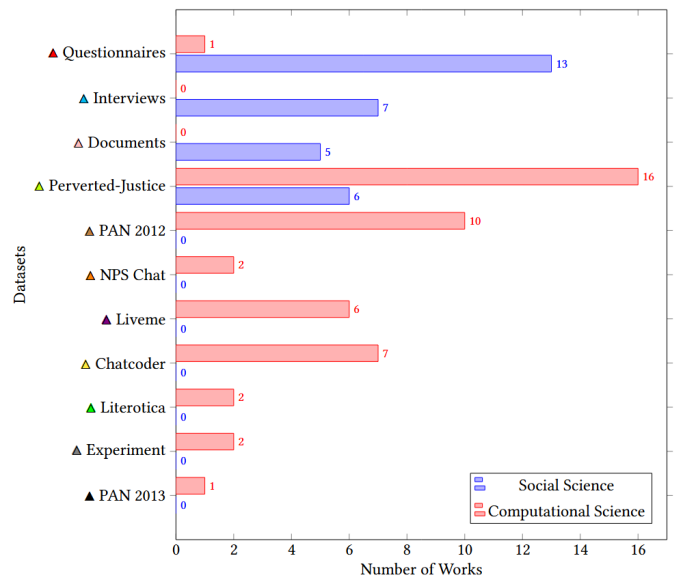
\begin{wrapfigure}{r}{0.6\textwidth}
\vspace{-5mm}
\centering
\resizebox{\linewidth}{!}{
\begin{tikzpicture}
\pgfplotsset{compat=1.16}
\begin{axis}[
    xbar,
bar width=5pt,
    y dir=reverse,
    xmin=0, xmax=20,
    height=9cm,
    width=0.53\textwidth,
    bar width=7pt,
    enlarge y limits=0.05,
    xlabel={Number of Works},
    ylabel={Datasets},
    ytick={2009,2010,2011,2012,2013,2014,2015,2016,2017,2018,2019},
    yticklabel style={font=\scriptsize, anchor=east, align=right},
    xticklabel style={font=\scriptsize},
    yticklabels={
        \textcolor{red}{$\blacktriangle$} \textbf{Questionnaires},
        \textcolor{cyan}{$\blacktriangle$} \textbf{Interviews},
        \textcolor{pink}{$\blacktriangle$} \textbf{Documents},
        \textcolor{lime}{$\blacktriangle$} \textbf{Perverted-Justice},
        \textcolor{brown}{$\blacktriangle$} \textbf{PAN 2012},
        \textcolor{orange}{$\blacktriangle$} \textbf{NPS Chat},
        \textcolor{violet}{$\blacktriangle$} \textbf{Liveme},
        \textcolor{yellow}{$\blacktriangle$} \textbf{Chatcoder},
        \textcolor{green}{$\blacktriangle$} \textbf{Literotica},
        \textcolor{gray}{$\blacktriangle$} \textbf{Experiment},
        \textcolor{black}{$\blacktriangle$} \textbf{PAN 2013}
    },
    nodes near coords,
    nodes near coords align={horizontal},
    legend style={
    font=\scriptsize,
    at={(0.98,0.02)},
    anchor=south east,
    legend columns=1,
    draw=white,
    fill=white,
    fill opacity=0.85,
    text opacity=1
},
legend image code/.code={
    \draw[#1, fill=#1] (0cm,-0.08cm) rectangle (0.4cm,0.08cm);
},
    legend entries={Social Sciences, Computational Sciences}
]
\addplot[fill=blue!30, draw=blue, nodes near coords style={font=\scriptsize, color=blue}]
    coordinates {(18,2009)(8,2010)(5,2011)(6,2012)(0,2013)(0,2014)(0,2015)(0,2016)(0,2017)(0,2018)(0,2019)};
\addplot[fill=red!30, draw=red, nodes near coords style={font=\scriptsize, color=red}]
    coordinates {(3,2009)(1,2010)(0,2011)(16,2012)(10,2013)(2,2014)(6,2015)(7,2016)(2,2017)(4,2018)(1,2019)};
\end{axis}
\end{tikzpicture}
}
\vspace{-8mm}
\caption{Overall trends of datasets in the reviewed cybergrooming research.}
\label{fig:datasets}
\vspace{-3mm}
\end{wrapfigure}

Computational science, in contrast, focuses on \textbf{\em machine learning (ML)} and \textbf{\em natural language processing (NLP)} to detect grooming behaviors and analyze communication patterns. ML models efficiently process vast datasets, identifying patterns and anomalies indicative of grooming activity, reducing the burden on human moderators. Techniques such as \textbf{\em sentiment analysis} assess emotional cues, \textbf{\em chatbot simulations} replicate grooming scenarios for training and prevention, and \textbf{\em automated detection algorithms} provide real-time monitoring. NLP enables the identification of subtle linguistic cues, such as persuasion and manipulation tactics. The methods used in cybergrooming research are summarized in \new{Tables~\ref{tab:evaluation-methods-social} and \ref{tab:evaluation-methods-computational} in the appendices,} %\new{Appendix~\ref{appendix:EvalMethodSS} and Appendix~\ref{appendix:evalmethodcs}}, 
with Fig.~\ref{fig:datasets} providing an overview of the datasets employed.

Social science research provides a detailed understanding of the \textbf{\em personal and psychological aspects} of grooming, shaping targeted prevention programs, and uncovering social dynamics that increase vulnerability. These studies support \textbf{\em policy development and education initiatives} by informing intervention strategies tailored to groomers' and victims' behaviors and motivations. \\ \vspace{-4mm}
\new{Recent prevention efforts operationalize parental education and awareness through evaluated interventions, including multilingual web platforms supporting parents, educators, and children in recognizing and responding to online grooming~\cite{kupcova2024innovative}. These approaches align with empirical evidence linking parental attitudes and protective behaviors to adolescents’ online risk exposure~\cite{lundrigan2025relationship}.}

Computational research contributes by developing \textbf{\em scalable, real-time monitoring and detection systems} that process vast amounts of online interactions, identifying grooming behaviors and supporting law enforcement efforts. The integration of \textbf{\em machine learning models and NLP tools} enhances detection accuracy and efficiency, enabling timely intervention. While social science research refines our understanding of grooming behaviors, computational methods operationalize this knowledge into technological solutions that can be deployed at scale.

\begin{wrapfigure}{r}{0.6\textwidth}
\vspace{-3mm}
\centering
\begin{tikzpicture}
\begin{axis}[
    xbar=0.7,
    y dir=reverse,
    xmin=0, xmax=20,
    height=7cm,
    width=0.48\textwidth,
    bar width=6pt,
    enlarge y limits=0.12,
    xlabel={Number of Works},
    ylabel={Limitations},
    ytick={1,2,3,4,5,6,7},
    tick label style={font=\scriptsize},   % 안전하게
    yticklabel style={anchor=east},        % 안전하게
    xticklabel style={font=\scriptsize},
    yticklabels={
        \textbf{Data Issue},
        \textbf{Generalizability},
        \textbf{Sexual Minority},
        \textbf{Cultural Difference},
        \textbf{Interdisciplinary},
        \textbf{Computing Power},
        \textbf{Raising Awareness}
    },
    nodes near coords,
    nodes near coords align={horizontal},
    every node near coord/.append style={font=\tiny}, 
    legend style={font=\scriptsize, at={(0.7,0.02)}, anchor=south, legend columns=1, draw=none},
legend image code/.code={
    \draw[#1, fill=#1] (0cm,-0.08cm) rectangle (0.4cm,0.08cm);
},    legend entries={Social Sciences, Computational Sciences}
]
\addplot[fill=blue!30, draw=blue, nodes near coords style={color=blue}]
    coordinates {(18,1)(17,2)(3,3)(3,4)(1,5)(0,6)(4,7)};
\addplot[fill=red!30, draw=red, nodes near coords style={color=red}]
    coordinates {(13,1)(11,2)(0,3)(6,4)(6,5)(3,6)(3,7)};
\end{axis}
\end{tikzpicture}
\vspace{-2mm}
\caption{Key limitations in the reviewed cybergrooming research.}
\label{fig:limitations}
\vspace{-3mm}
\end{wrapfigure}
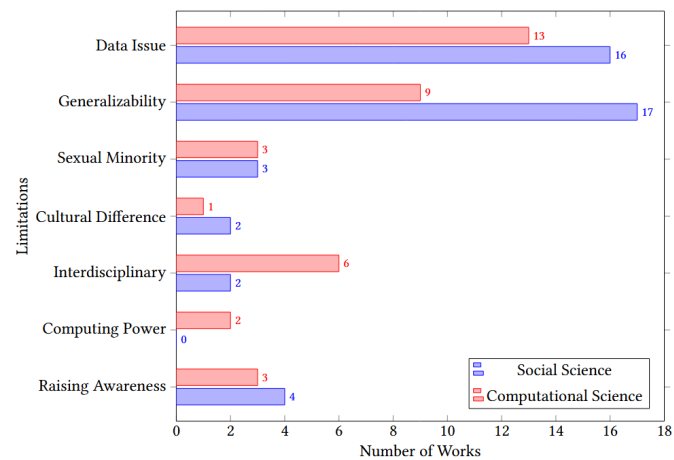
\textbf{RQ3.} \textbf{\em What limitations exist in the methodologies used across different disciplines?}

Social science research is often constrained by \textbf{\em reliance on self-reported data and small samples}, introducing biases and limiting generalizability. While qualitative methods offer rich insights, they are resource-intensive and susceptible to researchers' biases. Quantitative methods enable large-scale data collection. However, they depend on self-reports and require extensive resources for longitudinal studies.

Computational science faces challenges in \textbf{\em dataset availability, bias, and model limitations}. Effective training requires \textbf{\em extensive, diverse datasets}, often constrained by privacy limits. ML models may inherit biases, yielding skewed results and risking \textbf{\em overfitting}, performing well on training but poorly on unseen or cross-cultural data. Algorithms also struggle with nuanced interactions such as sarcasm, idioms, and contextual deception. Risks of \textbf{\em false positives and false negatives} further complicate deployment. Fig.~\ref{fig:limitations} summarizes these issues.

\textbf{RQ4.} \textbf{\em How do social and computational studies on cybergrooming differ, and what interdisciplinary insights can enhance the mitigation of cybergrooming risks?}

Cybergrooming studies in social and computational sciences differ in \textbf{\em focus and methodology}. Social sciences examine psychological, behavioral, and societal dimensions through interviews and surveys, yielding insights into victim and perpetrator profiles. Computational sciences rely on \textbf{\em data-driven techniques}, including ML and NLP, for automated detection of grooming.  Interdisciplinary collaboration strengthens mitigation strategies by integrating \textbf{\em social science insights into computational models}. Social science research clarifies grooming tactics, refining detection algorithms by improving sensitivity to subtle linguistic cues. Conversely, computational approaches provide \textbf{\em scalable tools for large-scale analysis}, enabling broader monitoring and intervention. A combined approach fosters \textbf{\em ethical, culturally aware solutions}, supporting adaptive detection systems that evolve with changing grooming tactics.

\textbf{RQ5.} \textbf{\em What gaps and challenges remain in cybergrooming research, and what future directions are needed?}

A critical gap is the need for \textbf{\em systematic integration of social and computational approaches} to leverage the strengths of both fields. While social sciences contribute behavioral insights, computational methods provide \textbf{\em scalable detection capabilities}. Groomers continually adapt their strategies, requiring \textbf{\em flexible, evolving detection models} that remain effective over time. Additionally, the \textbf{\em global nature of online interactions} highlights the importance of culturally inclusive solutions that can be applied across diverse contexts.

Future research should focus on \textbf{\em interdisciplinary frameworks} that facilitate collaboration between social scientists and computational researchers. Expanding \textbf{\em data availability and sharing mechanisms}, while ensuring privacy and ethical compliance, is essential for improving model training and validation. Developing \textbf{\em culturally adaptive machine learning models} that account for linguistic diversity will enhance detection accuracy across global digital environments. Furthermore, \textbf{\em community engagement and stakeholder involvement} in prevention strategy development can lead to more practical, widely accepted solutions. Lastly, integrating \textbf{\em multimodal data analysis}, including visual and auditory cues alongside text-based detection, can provide a more comprehensive understanding of grooming behaviors.

By bridging the gap between social science insights and computational scalability, future cybergrooming research can develop more effective, ethical, and adaptable mitigation strategies.

\subsection{Differences in the Used Datasets in Social and Computational Science Cybergrooming Research}
\label{sec:datasets_twofields}

We provided details of the datasets used in cybergrooming research, summarized in Tables~\ref{tab:datasets-social} and~\ref{tab:datasets-computational}. Datasets in both categories combine quantitative and qualitative data to offer comprehensive insights into online interaction patterns, addressing the behaviors of both predators and victims. However, they differ in several key respects. 

First, in terms of \textbf{what data to collect}, social science emphasizes personal surveys and self-reported experiences to capture victim perspectives and prevalence rates (e.g., Childhood Online Sexual Abuse in the US), whereas computational science relies on automated collection from online platforms, focusing on behavioral signals and pattern recognition (e.g., chat logs from Literotica and NPS Chat).  Second, regarding \textbf{data usage}, social science datasets provide detailed individual responses that support in-depth psychological and sociological analysis of grooming risks and victim vulnerability, while computational science datasets are structured for large-scale processing, enabling broad pattern detection, real-time risk assessment, and machine learning-driven classification.  Third, with respect to \textbf{how data are collected}, social science depends on direct engagement with participants through surveys and interviews to ensure contextual depth, whereas computational science employs automated techniques such as data mining, text analysis, and machine learning to extract insights from large-scale digital communication data.

To further distinguish these datasets, we propose a taxonomy based on their \textbf{\em scope and purpose}. First, \textbf{general-purpose datasets} capture broad interaction patterns and support diverse research questions; examples include the NPS Chat dataset for linguistic analysis and PAN12, which provides predatory and non-predatory conversations for cybercrime research. Second, \textbf{domain-specific datasets} target particular aspects of cybergrooming, such as law enforcement cases or psychological assessments; for example, the Childhood Online Sexual Abuse in the US dataset examines victimization experiences, while the Perverted Justice dataset contains decoy–predator chat logs used for detection studies. Third, \textbf{real-time datasets} are continuously updated to reflect ongoing online interactions; the LiveMe SLSS dataset, for instance, collects live-stream chat data to study grooming behaviors in real-time settings.

Understanding these dataset categories provides a clearer framework for selecting appropriate datasets based on research objectives. Social science research benefits from domain-specific and general-purpose datasets to study human behavior and victimization trends, while computational science relies heavily on large-scale general-purpose and real-time datasets to train and refine machine learning models. The integration of insights from both fields can enhance cybergrooming research by balancing behavioral depth with automated scalability.

\subsection{\new{AI-based Approaches in Cybergrooming Research: Progress and Limitations}}

\new{Across the literature, artificial intelligence is a core methodology in computational cybergrooming research, primarily supporting automated detection and large-scale analysis of online interactions~\cite{kim2020analysis, bours2019detection, rezaee2023detecting}. Most studies apply supervised or deep learning models to classify grooming-related content, identify risk signals, or infer grooming stages from annotated text~\cite{cano2014detecting, kim2023ai, farag2023enhanced}, with effectiveness largely demonstrated in controlled settings.}

\new{Despite these advances, existing AI systems remain largely detection-centric, prioritizing classification accuracy while offering limited support for intervention, prevention, or real-time decision-making~\cite{rita2021chatbot, Wang21-eancs}. Consequently, AI is often positioned as a retrospective analytical tool rather than an active protective mechanism, with adaptive or agent-based approaches still facing challenges related to robustness, false positives, and ethical accountability~\cite{daish2024towards, mujtaba2025edgeaiguard}.}

\new{Moreover, many models rely on static datasets and message-level analysis, limiting their ability to capture the longitudinal and interactional dynamics emphasized in social science research~\cite{fauzi2020ensemble, waezi2024osprey, fauzi2023identifying, milon2022take}. This gap constrains sensitivity to contextual shifts and relational progression. In addition, system effectiveness depends not only on model performance but also on integration into human-centered workflows, where challenges in interpretability, usability, and practitioner alignment remain unresolved~\cite{Munoz_Isaza_Castillo_2021, guo2023text}.}

\section{Conclusions \& Future Research Directions}

\new{This study presented a systematic interdisciplinary review of cybergrooming research across social and computational domains, synthesizing existing work, identifying structural limitations, and outlining future research directions. Using the PRISMA methodology~\cite{page2021prisma}, we ensured transparent and reproducible analysis of methods, datasets, evaluation practices, and research trends specific to cybergrooming.}

\new{Key findings indicate that social science research provides deep qualitative insights but is resource-intensive, while computational approaches enable scalable detection yet remain constrained by data quality, annotation bias, class imbalance, and detection-centric system designs. Across both domains, limited longitudinal modeling and weak integration into practitioner workflows persist, highlighting the need for contextualized and interpretable systems.}

\new{Future research should emphasize interdisciplinary integration of behavioral theory and AI, including culturally adaptive models, longitudinal and privacy-preserving learning, and closer collaboration among researchers, practitioners, and policymakers to advance ethical and human-centered cybergrooming prevention.}

\begin{acks}
This work is partially supported by the Commonwealth Cyber Initiative (CCI) Southwest Virginia (SWVA) Cybersecurity Research Program and the National Science Foundation (NSF) Secure and Trustworthy Cyberspace (SaTC) program under grants 2330940 and 2330941.
\end{acks}

%\nocite{*}
\bibliographystyle{ACM-Reference-Format}
\bibliography{ref}

%\clearpage
\appendix
\section*{APPENDICES}
%\section{Evaluation Methods in Social Science Research} \label{appendix:EvalMethodSS}
\begin{table}[htbp]
\footnotesize
    \centering
    \caption{\new{Evaluation methods used in the reviewed social science studies on cybergrooming. Studies employing multiple evaluation methods are listed under each relevant category.}}
    \label{tab:evaluation-methods-social}    
    \vspace{-3mm}
    \begin{tabular}{p{2.5cm} p{8.5cm} p{3cm}}
%    \hline
 %   \multicolumn{3}{c}{{\textbf{\gray Social Sciences}}} \\
    \hline
        \multicolumn{1}{c}{\textbf{Evaluation Method}} & \multicolumn{1}{c}{\textbf{Description}} & \multicolumn{1}{c}{\textbf{Research Work}} \\
    \hline
        Questionnaires & Collect quantitative data through structured questions; repeated to assess short- and long-term effects. Used to study online behaviors, victimization, demographics, personality, and other factors. & \cite{almeida2024online, Calvete_Orue_GamezGuadi_2022, Finkelhor22-online-sexual-offense, gamez2023stability, HERNANDEZ2021106569, wachs2020routine, Wachs2012, Weingraber2020, sani_social_2021, gamez-guadix_construction_2018, wachs2016cross, gamez2018persuasion, schoeps2020risk, carmo2023knowledge, villacampa2017online, almeida2025prevalence, lundrigan2025relationship, melo2025grooming} \\
    \hline
        Document Analysis & Qualitative review of documents to gather insights on research topics. Analyzed legal texts, books, journals, and other documents on cybergrooming and child protection laws. & \cite{Anggraeny2023, Fernandes_et_al_2023, ringenberg2022prepost, de_santisteban_progression_2018, choo_responding_2009, dolev-cohen_parental_2024, craven_current_2007, shannon2008sweden, seymour2021discursive} \\
    \hline
        Interviews & Collect in-depth information via conversations. Explored themes like parental awareness, criminal motives, and social media risks. & \cite{Anggraeny2023, Dorasamy_etal_2021, Fernandes_et_al_2023, de_santisteban_progression_2018, kloess2019case, Chiu2022onlinegrooming, christiansen2024hidden, khotimah2024child, reneses2024he} \\
    \hline
        Comparative Analysis & Compare items to highlight similarities/differences. Analyzed phases of cybergrooming investigations. & \cite{Fernandes_et_al_2023, soldino_criminological_2024, broome_psycho-linguistic_2020} \\
    \hline
        Deductive Technique & Draw specific conclusions from general principles. Investigated stages of cyber child grooming. & \cite{Anggraeny2023, soldino_criminological_2024} \\
    \hline
        Data Visualization Techniques & Methods for visual representation of data. Treemap Diagrams and Word Clouds were used to highlight key terms and visualize word frequencies related to cybergrooming. & \cite{Dorasamy_etal_2021} \\
    \hline
        Statistical Analysis & Various methods to analyze and interpret quantitative data, including statistical software. ANCOVA, ANOVA, chi-square test, t-test, logistic regression, multiple linear regression analysis, and Pearson’s correlation used for hypothesis testing. & \cite{almeida2024online, Finkelhor22-online-sexual-offense, gamez2023stability, HERNANDEZ2021106569, wachs2020routine, Wachs2012, Weingraber2020, Calvete_Orue_GamezGuadi_2022, soldino_criminological_2024, gamez-guadix_construction_2018, craven_current_2007,wachs2016cross, van2016behavioural, gamez2018persuasion, schoeps2020risk, aktu2024self, resett2025prediction} \\
    \hline
        Other Tools & Tools to interpret social science data and phenomena. Computer-Mediated Discourse Analysis, Structural Equation Modeling, thematic analysis, discourse analysis, and generic qualitative inquiry. & \cite{Lorenzo-Dus_Izura_2017, Dorasamy_etal_2021, HERNANDEZ2021106569, broome_psycho-linguistic_2020, grant_assuming_2016, broome_investigation_2024, schoeps2020risk, carmo2023knowledge, evans2025corpus, lorenzo5280035not} \\
    \hline
    \end{tabular}
    \vspace{-5mm}
\end{table}

%\clearpage
%\section{Evaluation Metrics in Social Science Research} \label{appendix:evalmetricsss}
\begin{table}[htbp]
\footnotesize
    \centering
    \caption{Evaluation Metrics in the Reviewed Social Science Research Works on Cybergrooming}
    \label{tab:evaluation-metrics-social}    
\vspace{-3mm}    
\begin{tabular}{p{3.5cm} p{7.5cm} p{3cm}} 
    \hline
   % \multicolumn{3}{c}{\textbf{\gray Social Sciences}} \\
   % \hline
        \multicolumn{1}{c}{\textbf{Evaluation Metric}} & \multicolumn{1}{c}{\textbf{Description}} & \multicolumn{1}{c}{\textbf{Research Work}} \\
    \hline
        Chi-square Statistics & An assessment of variable association, with higher values indicating greater significance. & \cite{Finkelhor22-online-sexual-offense, gamez2023stability, HERNANDEZ2021106569, wachs2020routine, Wachs2012, Weingraber2020} \\
    \hline
        Frequency & A measure of theme frequency, identifying prevalent patterns. & \cite{Anggraeny2023, Dorasamy_etal_2021, Fernandes_et_al_2023, Lorenzo-Dus_Izura_2017, broome_psycho-linguistic_2020, sani_social_2021, broome_investigation_2024} \\
    \hline
        $t$-statistics & An evaluation of the statistical significance of differences between group means. & \cite{Calvete_Orue_GamezGuadi_2022, HERNANDEZ2021106569, wachs2020routine, Weingraber2020, soldino_criminological_2024, sani_social_2021, broome_investigation_2024} \\
    \hline
        Comparative Fit Index (CFI) & An indicator of model fit, with values closer to one suggesting a better fit. & \cite{HERNANDEZ2021106569, wachs2020routine, gamez-guadix_construction_2018, schoeps2020risk} \\
    \hline
        Composite Reliability Coefficient & Testing the internal consistency in Structural Equation Modeling models. & \cite{HERNANDEZ2021106569, wachs2020routine} \\
    \hline
        Confidence Interval & A range that likely contains the true parameter, indicating the level of precision. & \cite{wachs2020routine, Wachs2012} \\
    \hline
        Correlation Coefficients & Shows direction and strength of relationships between variables. & \cite{Calvete_Orue_GamezGuadi_2022, Weingraber2020} \\
    \hline
        Cronbach's Alpha & Measuring the internal consistency of responses. & \cite{HERNANDEZ2021106569, Wachs2012, gamez-guadix_construction_2018, schoeps2020risk} \\
    \hline
        Prevalence Rate & Measuring the proportion of population meeting a condition, indicating a phenomenon extent. & \cite{Finkelhor22-online-sexual-offense, Wachs2012} \\
    \hline
        RMSEA & Evaluating model fit; lower values suggest better fit. & \cite{HERNANDEZ2021106569, wachs2020routine, schoeps2020risk} \\
    \hline
        Standard Error & Estimating the variability of sample statistics, indicating precision. & \cite{Finkelhor22-online-sexual-offense, wachs2020routine} \\
    \hline
        Tucker-Lewis Index (TLI) & Assessing model fit; values close to one indicate good fit. & \cite{HERNANDEZ2021106569, wachs2020routine} \\
    \hline
        Average Variance Extracted (AVE) & Evaluating measurement model quality in Structural Equation Modeling. & \cite{HERNANDEZ2021106569} \\
    \hline
        Contingency Coefficient & Measuring association strength between two categorical variables. & \cite{Weingraber2020} \\
    \hline
        $f$-statistics & Evaluating overall significance of a regression model. & \cite{Wachs2012} \\
    \hline
        Guttman's Lambda 6 (G6) & Assessing data consistency; values near one indicate reliability. & \cite{Wachs2012} \\
    \hline
        Odds Ratio & Measuring association between two events; indicates their relationship. & \cite{Wachs2012} \\
    \hline
        Partial Eta Squared & Measuring the effect size and variance explained by a model. & \cite{Wachs2012, broome_psycho-linguistic_2020, sani_social_2021, broome_investigation_2024} \\
    \hline
        Regression Coefficients & Evaluating strength and direction of relationships between variables. & \cite{Calvete_Orue_GamezGuadi_2022} \\
    \hline
    \end{tabular}
%    \vspace{-5mm}
\end{table}

%\clearpage
%\section{Evaluation Methods in Computational Science Research} \label{appendix:evalmethodcs}
\begin{table}[htbp]
%\vspace{-4mm}
\footnotesize
    \centering
    \caption{Evaluation Methods in Computational Science Studies on Cybergrooming}
    \label{tab:evaluation-methods-computational}
\vspace{-3mm}
    \begin{tabular}{p{3.5cm} p{8cm} p{2.5cm}}
%    \hline
%    \multicolumn{3}{c}{\textbf{\gray Computational Sciences}} \\
    \hline
         \multicolumn{1}{c}{\textbf{Evaluation Method}} &  \multicolumn{1}{c}{\textbf{Description and Application}} & \multicolumn{1}{c}{\textbf{Research Work}} \\
    \hline
        Natural Language Processing (NLP) & Techniques for processing and analyzing natural language data, including tokenization, parsing, and sentiment analysis to be applied in various studies on text data & \cite{amer2021detection, gupta2012characterizing, bogdanova2014exploring, cano2014detecting, Borj_Bours_2019, Munoz_Isaza_Castillo_2021, Lykousas_Patsakis_2022, Eilifsen_Shrestha_Bours_2023, guo2023text, isaza2022classifying, anderson2019intelligent, cook2023protecting, farag2023enhanced, ringenberg2024assessing} \\
    \hline
        Support Vector Machine (SVM) & A supervised learning model for classification and regression, identifying hyperplanes that separate classes in feature space & \cite{bogdanova2014exploring, Gunawan16, Borj_Bours_2019, sulaiman2019classification, anderson2019intelligent, rezaee2023detecting} \\
    \hline
        Convolutional Neural Networks (CNNs) & Deep neural networks primarily for visual imagery, applied to text for feature extraction and classification & \cite{Munoz_Isaza_Castillo_2021, isaza2022classifying, guo23-is} \\
    \hline
        LIWC (Linguistic Inquiry and Word Count) & A text analysis tool measuring word category usage across texts, useful for psychological and social research & \cite{gupta2012characterizing, cano2014detecting, kim2023ai} \\
    \hline
        Naïve Bayes & A probabilistic classifier based on Bayes' theorem to be effective for text classification tasks due to its assumption of feature independence & \cite{Gunawan16, sulaiman2019classification} \\
    \hline
        Agile Software Development & Methodologies for efficient project management in software development, emphasizing flexibility and customer feedback & \cite{rita2021chatbot} \\
    \hline
        Large Language Models (LLMs) & Advanced transformer models such as BERT, T5, RoBERTa, and BERT for NLP tasks, known for their contextual understanding of language & \cite{Wang21-eancs, rezaee2023detecting, prosser2024helpful, mujtaba2025edgeaiguard} \\
    \hline
        Binary Logistic Regression & A model predicting binary outcomes by estimating probabilities, useful for classifying observations into two categories & \cite{pranoto2015logistic} \\
    \hline
        $k$-Nearest Neighbors & An instance-based learning algorithm classifying data by comparing it to similar data points in the training set & \cite{Gunawan16} \\
    \hline
        Linear Discriminant Analysis & A method finding linear combinations of features that best separate classes, used in pattern recognition tasks & \cite{Lykousas_Patsakis_2022} \\
    \hline
        Psycho-linguistic profiling & Analysis of text to understand psychological states and traits, often using tools like LIWC & \cite{gupta2012characterizing} \\
    \hline
        Random Forest & Combining multiple decision trees for accurate predictions while reducing overfitting & \cite{Gunawan16} \\
    \hline
        Recurrent Neural Networks & Neural networks with directed connections capable of handling temporal sequences for dynamic behavior & \cite{kim2020analysis, guo23-is} \\
    \hline
        Universal Sentence Encoder & Using LSTM to generate message vectors, preserving word relationships for conversational context classification & \cite{kim2020analysis} \\
    \hline
        Word2Vec & Converting words into vector representations to capture semantic similarities, aiding in classification and clustering & \cite{Munoz_Isaza_Castillo_2021} \\
    \hline
        Temporal Concept Analysis (TCA) & TCA with Temporal Relational Semantic Systems, conceptual scaling, and nested line diagrams to analyze chat conversations, providing insights into the temporal dynamics and context of conversations for potential risk detection & \cite{elzinga2012analyzing} \\
    \hline
        Game Theory & Proposing game-theoretic approaches to model and simulate threat scenarios & \cite{kim2023ai} \\
    \hline
    \end{tabular}
    \vspace{-5mm}
\end{table}

%\clearpage
%\vspace{-3mm}
%\section{Evaluation Metrics in Computational Science Research} \label{appendix:evalmetricscs}
\begin{table}[htbp]
\footnotesize
    \centering
    \caption{Evaluation Metrics in Computational Science Studies on Cybergrooming}
    \label{tab:evaluation-metrics-computational}
\vspace{-3mm}
    
    \begin{tabular}{p{3.5cm} p{7.5cm} p{3cm}}
%    \hline
%    \multicolumn{3}{c}{\textbf{\gray Computational Sciences}} \\
    \hline
        \multicolumn{1}{c}{\textbf{Evaluation Metric}} & \multicolumn{1}{c}{\textbf{Description and Application}} & \multicolumn{1}{c}{\textbf{Research Work}} \\
    \hline
        Precision & Proportion of true positives among all positive predictions, helping identify relevant cases in classification tasks. & \cite{cano2014detecting, Ashcroft_Kaati_Meyer_2015, Borj_Bours_2019, bours2019detection, fauzi2020ensemble, kim2020analysis, Munoz_Isaza_Castillo_2021, Eilifsen_Shrestha_Bours_2023, amer2021detection, isaza2022classifying, pranoto2015logistic, sulaiman2019classification, cook2023protecting, rezaee2023detecting} \\
    \hline
        Accuracy & Measures the proportion of correctly identified cases (both positives and negatives) out of the total cases examined, indicating overall model performance. & \cite{bogdanova2014exploring, Ashcroft_Kaati_Meyer_2015, Gunawan16, Borj_Bours_2019, Munoz_Isaza_Castillo_2021, fauzi2020ensemble, Eilifsen_Shrestha_Bours_2023, amer2021detection, isaza2022classifying, pranoto2015logistic, sulaiman2019classification, cook2023protecting, rezaee2023detecting} \\
    \hline
        Recall & Proportion of actual positives correctly identified, demonstrating the model’s ability to capture all relevant cases. & \cite{cano2014detecting, Ashcroft_Kaati_Meyer_2015, Borj_Bours_2019, bours2019detection, fauzi2020ensemble, kim2020analysis, Munoz_Isaza_Castillo_2021, Eilifsen_Shrestha_Bours_2023, isaza2022classifying, pranoto2015logistic, sulaiman2019classification, cook2023protecting, rezaee2023detecting} \\
    \hline
        F Score & Combines precision and recall into a single metric, with the F1 score balancing them equally and the $F\beta$ score adjusting this balance based on specific needs. & \cite{cano2014detecting, Borj_Bours_2019, bours2019detection, kim2020analysis, fauzi2020ensemble, Munoz_Isaza_Castillo_2021, Eilifsen_Shrestha_Bours_2023, amer2021detection, isaza2022classifying, sulaiman2019classification, cook2023protecting, rezaee2023detecting} \\
    \hline
        ROC Curve and AUROC & The ROC curve plots true positive rates against false positive rates; AUROC represents the area under the curve, indicating the model's class distinction ability. & \cite{Munoz_Isaza_Castillo_2021, Eilifsen_Shrestha_Bours_2023, isaza2022classifying, amer2021detection} \\
    \hline
        Confusion Matrix & A table summarizing the performance of a classification model by showing counts of true positives, true negatives, false positives, and false negatives. & \cite{pranoto2015logistic, amer2021detection, Ashcroft_Kaati_Meyer_2015} \\
    \hline
        BLEU, ROUGE, BERTScore & Metrics evaluating machine-generated text quality, focusing on n-gram precision, recall, and contextual similarity. & \cite{Wang21-eancs} \\
    \hline
        Specificity & Measures the proportion of true negatives correctly identified by the model, relevant in classification tasks. & \cite{Munoz_Isaza_Castillo_2021} \\
    \hline
        Youden's Index & A single metric that balances sensitivity and specificity, aiding in the evaluation of diagnostic tests. & \cite{Eilifsen_Shrestha_Bours_2023} \\
    \hline
        Human Evaluation & Involves annotators comparing machine-generated content with human-created content to assess quality and naturalness. & \cite{Wang21-eancs,kupcova2024innovative, roumelioti2025leveraging, daish2024towards, mateo2025groombuster} \\
    \hline
        Topic Coherence & Measures coherence of topics generated by modeling algorithms, indicating alignment with human understanding. & \cite{Lykousas_Patsakis_2022} \\
    \hline
    \end{tabular}
    \vspace{-5mm}
\end{table}

\end{document}